\documentclass[a4paper,12pt]{article}
\pdfoutput=1 

\usepackage{jheppub} 

\usepackage[T1]{fontenc} 

\usepackage{lscape}
\usepackage{tikz}
\usepackage{array}
\usepackage{hyperref}
\usepackage{amsfonts}
\usepackage{amssymb}
\usepackage{amsmath, amsthm}
\usepackage{bbm} 
\usepackage{graphicx}
\usepackage{caption}
\usepackage{mathtools}
\usepackage{imakeidx}
\usepackage[utf8]{inputenc}
\usepackage[T1]{fontenc}
\usepackage{booktabs}
\usepackage{multirow}
\makeindex

\newcommand{\te}{\textrm{e}}

\renewcommand{\=}{\,= \,}

\newcommand\G{\Gamma}

\newcommand{\z}{\zeta}

\newcommand\bi{\begin{itemize}}
\newcommand\ei{\end{itemize}}

\newcommand\bspl{\begin{split}}
\newcommand\espl{\end{split}}

\newcommand{\susyL}[1]{\underset{\text{SUSY Locus}}{\longrightarrow}}
\newcommand{\be}{\begin{equation}}
\newcommand{\ee}{\end{equation}}
\newcommand{\bea}{\begin{eqnarray}}
\newcommand{\eea}{\end{eqnarray}}

\renewcommand{\=}{\,= \,}

\newtheorem{theorem}{Observation}[section]

\theoremstyle{definition}
\newtheorem{definition}{Definition}[section]

\theoremstyle{definition}
\newtheorem{property}{Property}[section]

\begin{document}

\title{Generalized Bethe expansions of superconformal indices}

\author{\,Alejandro Cabo-Bizet\,$^a$ and  Wei Li\,$^b$}

\affiliation[a]{\,\,Università del Salento, Dipartimento di Matematica e Fisica Ennio De Giorgi, and I.N.F.N. - sezione di Lecce, Via Arnesano, I-73100 Lecce, Italy}
\affiliation[b]{\,\,Institute of Theoretical Physics, Chinese Academy of Sciences, 100190 Beijing, P.R.\ China}

\emailAdd{acbizet@gmail.com}
\emailAdd{weili@mail.itp.ac.cn}

\abstract{ 
We show the existence of infinitely many Bethe expansions for general four dimensional superconformal theories.
We then propose an analytic method to systematically obtain all the Bethe solutions, both isolated and continuous, for general superconformal theories.
In particular, we show that the contribution from the continuous manifold of solutions to the index comes from isolated points on this manifold. 
We check our proposals on $\mathcal{N}=4$ SYM.
For $U(3)$ or $SU(3)$, which are the first examples with continuous solutions, we demonstrate the non-trivial cancellation between the tachyon contributions from the previously known isolated solutions and those from the continuous solutions. 
}

\maketitle

\section{Introduction and summary}
\label{sec1}

Superconformal gauge theories  have proven to be a valuable playground for refining our understanding of non-perturbative effects in gauge theories and in holography.

Following two ground-breaking results in the context of string theory, namely the description of the microstates of certain asymptotically flat supersymmetric black hole as brane excitations~\cite{Strominger:1996sh} and the development of AdS/CFT~\cite{Maldacena:1997re}, it was naturally proposed that AdS black hole microstates should be understood as gauge-invariant states in a dual superconformal gauge theory~\cite{Witten:1998zw}.
In the supersymmetric context, the hope was to make this correspondence between bulk microstates and gauge-invariant boundary states precise, see e.g.~\cite{Kinney:2005ej,Berkooz:2006wc,Kim:2006he,Janik:2007pm,Grant:2008sk,Chang:2013fba} for notable early attempts.

This expectation was first 
confirmed by~\cite{Benini:2015eyy} for AdS$_4$/CFT$_3$, namely for the holographic duality between string theory in AdS$_4\times \mathbb{CP}^3$ and certain 3D $\mathcal{N}=6$ superconformal Chern-Simons-matter theory (the so-called ABJM theory) \cite{Aharony:2008ug}. 
Soon after, it was also established by~\cite{Cabo-Bizet:2018ehj,Choi:2018hmj, Benini:2018ywd} for AdS$_5$/CFT$_4$, namely for the duality between string theory in AdS$_5\times$S$^5$ and 4D $\mathcal{N}=4$ SYM. These latter results have since been further refined and extended to other $4$D superconformal theories~\cite{Cabo-Bizet:2019eaf,Aharony:2021zkr,Hosseini:2017mds,Kim:2019yrz,Cabo-Bizet:2019osg,Cassani:2019mms,Lezcano:2019pae,Lanir:2019abx,Bobev:2019zmz,ArabiArdehali:2019orz,Cabo-Bizet:2020nkr,Murthy:2020scj,Agarwal:2020zwm,Benini:2020gjh,Copetti:2020dil,Cabo-Bizet:2020ewf,Goldstein:2020yvj,Amariti:2021ubd,Cassani:2021fyv,ArabiArdehali:2021nsx,Jejjala:2021hlt,Ardehali:2021irq,Choi:2021rxi,Cabo-Bizet:2021plf,Cabo-Bizet:2021jar,Bobev:2021qxx,Bobev:2022bjm,Cassani:2022lrk,Jejjala:2022lrm,Cassani:2023vsa,Choi:2023tiq,Aharony:2024ntg}. 

There are various techniques employed in reaching these results. 
In this paper we study one of them: the Bethe ansatz formula for the superconformal index of 4D superconformal gauge theories~\cite{Benini:2018mlo,Closset:2017bse}.
 
The superconformal index 
$\mathcal{I}$ of a 4D superconformal gauge theory can be computed by an  integral over the gauge rapidities $z_i$~\cite{Romelsberger:2005eg,Kinney:2005ej,Dolan:2008qi}
\be
\mathcal{I}\,=\,\oint_{|z_i|=1} \prod_{i=1}^{N-1}\frac{\text{d}z_i}{2 \pi z_i}\cdot e^{-S[z,q]}\,,
\ee 
where $N$ is the rank of the gauge group.
After explicit integration, and for $|q|<1\,$, these integrals reduce to $q$-series of the form
\be
\sum_{n\,\geq\,0} a_n q^{J(n)}\,,  \qquad a_n\,\in\,\mathbb{Z}\,,
\ee
with $J(n)$ being some asymptotically growing positive function of $n\,$.
The integers $a_n$ count superconformal states at level $J(n)$ with $(-1)^F$ grading. 

\smallskip

The authors of~\cite{Closset:2017bse,Benini:2018mlo} proposed an alternative way to evaluate $4$D superconformal indices, a~\textit{Bethe expansion} of $\mathcal{I}$ of the form: 
\be\label{eq:BetheexpansionIntro}
\mathcal{I}\,=\,\sum_{z^{*}} e^{-S[z^{*},q]-\log H[z^{*},q]}\,.
\ee
This expansion is usually called the Bethe ansatz expansion.
The~$z^{*}$'s are solutions to a particular set of \textit{Bethe equations}~\cite{Benini:2018mlo} 
\be\label{eq:IntroBEq}
\{\mathcal{Q}_i[z,q]=1\}\,,
\ee
were $\mathcal{Q}_i$ are specific functions that we will call \textit{Bethe discontinuities}.\footnote{
They were also called ``Bethe operators" in~\cite{Benini:2018mlo}.} 
$H$ is a one-loop determinant contribution, which we will call the Jacobian of the \emph{Bethe solutions} $z^*\,$.
The functions $e^{S}$, $H$ and $\mathcal{Q}_i$ are related in a precise way that will be detailed in due time.

\smallskip

The advantage of the Bethe expansion~\eqref{eq:BetheexpansionIntro} over other representations of superconformal indices is that it has a very transparent interpretation in holography, with the power to exactly address non-perturbative questions such as background transition and backreaction of branes in geometry, etc. 
For example,  
at large charges of order $N^2\gg1$, where $N$ is the rank of the gauge group, the~$z^{*}$'s are expected to correspond to semiclassical gravitational saddle points, with the leading one corresponding to an Euclidean black hole saddle. 
This was shown in examples such as  4D $\mathcal{N}=4$ SYM~\cite{Cabo-Bizet:2018ehj,Aharony:2021zkr} and in various other 4D superconformal theories with known holographic duals.

The Bethe expansion~\eqref{eq:BetheexpansionIntro} is expected to give an \textit{exact} holographic dual representation of superconformal indices, even away from the semi-classical limit and in the full quantum regime.
From the perspective of the superconformal theory,  the $z^{*}$ classify {statistical phases} of the gauge theory~\cite{Choi:2018hmj,Benini:2018ywd,Cabo-Bizet:2019eaf,Aharony:2021zkr,Kim:2019yrz,Cabo-Bizet:2019osg,Lezcano:2019pae,Lanir:2019abx,ArabiArdehali:2019orz,Cabo-Bizet:2020nkr,Murthy:2020scj,Agarwal:2020zwm,Benini:2020gjh,Copetti:2020dil,Cabo-Bizet:2020ewf,Goldstein:2020yvj,Amariti:2021ubd,Jejjala:2021hlt,Ardehali:2021irq,Cabo-Bizet:2021plf,Cabo-Bizet:2021jar,Jejjala:2022lrm,Aharony:2024ntg}.
It is then natural to expect that, on the other hand, the $z^{*}$ classify statistical phases of the dual stringy formulation. 
(This perspective deserves further understanding as it might lead to a fresh viewpoint on holography.)

\medskip

Despite its clear strengths, the Bethe expansion~\eqref{eq:BetheexpansionIntro}, and its underlying physics,  have not been fully understood. 
For instance, even for the simplest 4D superconformal theoy with known holographic dual, namely the 4D $\mathcal{N}=4$ SYM with $U(N)$ gauge group, the formula has yet to be fully developed for $N\geq 3\,$.
And the complexity of the problem increases very fast with $N$.

\medskip

There are two main challenges on the way:
\begin{itemize}
\item[1.] We are lacking a method to determine \textit{all} Bethe solutions $z^{*}$ that contribute to the index, which we name ``contributing Bethe solutions". 
(Note that the Bethe equations for the superconformal indices often involve ratios of Jacobi theta functions, and hence are even more  difficult than the ordinary rational Bethe equations.)  
\item[2.] Some of the $z^{*}$ are not isolated points but (complex) manifolds.\footnote{
For example, the continuous solutions for $\mathcal{N}=4$ $U(3)$ SYM form a complex curve.}
For these ``continuous solutions" (with Jacobian $H=0$)~\cite{ArabiArdehali:2019orz}, it is not known how to evaluate their contributions to the index~\cite{ArabiArdehali:2019orz,Lezcano:2021qbj,Benini:2021ano}.
\end{itemize}

In this paper, we propose solutions to these two problems for generic superconformal gauge theories, and check them for  the simplest non-trivial case, the 4D $\mathcal{N}=4$ SYM with gauge group $U(3)$ (or $SU(3)$)\,.

\subsection{Summary of main results}

\paragraph{Generalized Bethe expansions}

We start by presenting, in Section~\ref{sec:2}, a broader perspective on the Bethe ansatz formulae of 4D superconformal indices, building upon the approach pioneered in~\cite{Benini:2018mlo}. 
\begin{enumerate}
\item We first define a set of \emph{shift operators} $\{\widehat{\mathcal{Q}}_j^{\underline{\underline{n}}}\}$ that rescale the gauge fugacities $\{z_i\}$. The information of $\{\widehat{\mathcal{Q}}^{\underline{\underline{n}}}_j\}$ is packaged in an integer \textit{periodicity matrix}~$\underline{\underline{n}}$, of dimension $(N-1)$, where $N$ is the rank of the gauge group, with 
$\det\underline{\underline{n}}\,\neq0\,$.\footnote{
The choice of Bethe operator used in~\cite{Benini:2021ano} (see between equations~(2.16) and~(2.23) there) corresponds to a particular choice of $\underline{\underline{n}}$, see equation~\eqref{eq:TranslationToBetheOperator} below.}
\item Under the action of the shift operators $\{\widehat{\mathcal{Q}}^{\underline{\underline{n}}}_j\}$, the integrand $I$ transforms quasi-periodically, acquiring the pre-factor $\{\mathcal{Q}^{\underline{\underline{n}}}_j\}$ which we call \textit{Bethe discontinuity}. 
We show that the infinitely many possible $\{\widehat{\mathcal{Q}}^{\underline{\underline{n}}}_j\}$ are in one-to-one correspondence with Bethe expansions of the superconformal index, which we call \textit{generalized Bethe formulae}, labeled by $\underline{\underline{n}}$.
\end{enumerate}
In summary, in this broader framework, for a given 4D superconformal theory, there exist an infinite number of Bethe expansions (in addition to the one derived in~\cite{Benini:2018mlo}), all of which give the same 
superconformal index $\mathcal{I}$. 

To demonstrate the power of these generalized Bethe formulae, we show (in  Section~\ref{sec:amb211}) that
they can be organized in orbits with common sets of Bethe solutions $z^{*}\,$, and that 
the part of the orbit that intersects the Bethe-ansatz formulae used in~\cite{Lezcano:2021qbj,Benini:2021ano} corresponds to the set of~$\underline{\underline{n}}$'s belonging to the subgroup of signed unipotent matrices of rank $N-1$, namely $\underline{\underline{n}}\,\in \,{U}^\ast(N-1,\mathbb{Z})\,\subset\,{GL}(N-1,\mathbb{Z})\,$.
\footnote{
It is possible that the orbit is even larger.} 
This resolves previous puzzles regarding the appropriate choice of Bethe operator and Bethe equations~\cite{Benini:2021ano}.

\paragraph{Contribution of continuous solutions}
We classify the solutions to the Bethe equations~\eqref{eq:IntroBEq} as follows:
\begin{enumerate}
\item \textit{Isolated} Bethe solutions. (For $\mathcal{N}=4$ SYM, they are also called \textit{topological} Bethe solutions.) 
\begin{enumerate}
\item \textit{Irreducible} (a.k.a.\ \textit{non-degenerate}) isolated Bethe solutions, 
with Jacobian $H\neq 0$.
\item \textit{Reducible} (a.k.a.\ \textit{degenerate}) isolated Bethe solutions, with Jacobian $H=0$.
\end{enumerate}
\item \textit{Continuous} Bethe solutions.\footnote{
Some of the isolated solutions can also lie on the manifold of the continuous solutions.} 
They have Jacobian $H= 0$.
(For $\mathcal{N}=4$ SYM, they are also called \textit{non-topological} Bethe solutions.) 
\end{enumerate}

Evaluating the contribution of continuous solutions to the index has been a standing problem~\cite{ArabiArdehali:2019orz,Lezcano:2021qbj,Benini:2021ano}.
In Section~\ref{subsec:MethodContinuous} we introduce a method  to compute the contributions from continuous Bethe solutions in generic superconformal indices. 
In particular we explain the mechanism by which they localize to isolated points. 

\paragraph{A solution-generating method}
In Section~\ref{eq:RecursiveMethod} we introduce a method to systematically identify the Bethe solutions $z^{*}$ that can potentially contribute to the superconformal index, for generic superconformal theories.

\paragraph{Application on SYM} 
In Section~\ref{sec:3} we apply the general formalism introduced in section~\ref{sec:2} to the superconformal index of $\mathcal{N}=4$ SYM, both at equal and different rapidities. 
We first focus on the case of equal rapidities $p=q$, and reproduce the Bethe discontinuity used in earlier work \cite{Benini:2021ano}. 

In Section~\ref{sec:3.4} we also introduce a novel choice of Bethe discontinuity for the cases with $p\neq q$\,, whose validity is then confirmed  a posteriori in Appendix~\ref{sec:PneqQ}.

\medskip

In Section~\ref{sec:4} we study the Bethe solutions for  $\mathcal{N}=4$ SYM with $U(N)$ or $SU(N)$ gauge group, and use them to demonstrate the solution classification.
We first consider the isolated roots, which are called topological solutions in this example. 
For the $U(N)$ SYM, the irreducible and reducible topological solutions are distinguished by their behavior under the cyclic symmetry $\mathbb{Z}_N$ of the Bethe equations~\eqref{eq:IntroBEq}.
While irreducible topological solutions can be written down uniformly for any $N$,\footnote{
Irreducible topological solutions are the Hong-Liu Bethe solutions~\cite{Hong:2018viz,Hosseini:2016cyf} and the fixed-points of~\cite{Cabo-Bizet:2019eaf,Cabo-Bizet:2020nkr,Cabo-Bizet:2020ewf}.} the reducible one need to be solved for each $N$. 
We give explicit solutions for $N=2,\dots, 5$.
(We also show, in Section~\ref{subsec:TopPNeqQ}, how to promote the topological solutions with equal rapidities $p=q$ to those with allowed non-equal pairs $p\neq q\,$.)

\smallskip

The continuous solutions (also called the non-topological solution in this case) break the cyclic symmetries $\mathbb{Z}_N$ (except for at the points that coincide with reducible topological solutions) and hence one cannot utilize the $\mathbb{Z}_N$ group to determine the continuous solutions.
However, once all the topological Bethe solutions (i.e.\ isolated roots) have been classified, we can use them to generate all the non-topological solutions (i.e.\ continuous solutions).
The new trick is to identify certain expansions where the Bethe equations become rational in the fugacities and hence analytically solvable. 

\smallskip

Then we show that  the contribution to the index from continuous solutions localizes to isolated points (which we call contributing non-topological solutions), which in particular never coincide with the reducible topological solutions. 

The classification of the solutions, together with their properties, for $U(N)$ or $SU(N)$ $\mathcal{N}=4$ SYM is summarized in table~\ref{tab:solutions}.

\smallskip

In Section~\ref{ref:AlgorithmSols} we introduce an alternative procedure that turns out to be optimal in determining the contributing solutions, bypassing the step of solving
Bethe equations.

\begin{table}[h!] \centering \begin{tabular}{lccc} \toprule \textbf{Bethe solutions} &
\textbf{Non-degenerate}& \textbf{Contributing} \\ \midrule 
Irreducible topological & 
\checkmark & 
\checkmark \\ 
Reducible topological & 
$\times$ &
$\times$ \\ 
Non-topological (a.k.a.\ continuous) & 
$\times$ & 
\checkmark
\\ \bottomrule 
\end{tabular} \caption{Classification of Bethe solutions for the superconformal index of $U(N)$ $\mathcal{N}=4$ SYM, their degeneration behavior, and their contributions to the index.
} \label{tab:solutions} \end{table}

Finally, in Section~\ref{sec:5} we check our proposals, specifically the approach introduced in Section~\ref{ref:AlgorithmSols}, by confirming that the tachyonic contributions coming from the known topological solutions~\cite{Lezcano:2021qbj,Benini:2021ano}, cancel against tachyonic contributions coming from continuous solutions in a rather non-trivial way. 
This is done for the first case where continuous solutions exist, namely the $U(3)$ SYM theory.

\section{Bethe expansions}\label{sec:2}

\subsection{Quasi-periodicities of superconformal indices}

In this paper, we are interested in computing 4D superconformal indices $\mathcal{I}\,$, which are integrals of the form
\begin{equation}
\label{eq:TheIntegral}
\mathcal{I}\,:=\,\oint_{\mathcal{C}_0} \prod_{i=1}^{N-1}\frac{dz_i}{2\pi\text{i}z_i}\,\cdot\, {I}[\underline{z}]\,,
\end{equation}
with the integrand\footnote{
The $\dots$ denotes the weights w.r.t.\ the remaining global charges if present, which we will largely ignore in this paper since they are not universally present for superconformal indices.}
\be\label{eq:IntegrandProd}
\begin{split}
I=I[\underline{z}]&\,{:=}\,\frac{\prod_{{F}}(1+p^{r_0 j_1[F]} q^{r_0 j_2[F]} \,\ldots 
\,\underline{z}^{\underline{\rho}[F]}
)}{\prod_{{B}}(1+p^{r_0 j_1[B]} q^{r_0 j_2[B]}  
\,\ldots 
\,\underline{z}^{\underline{\rho}[B]}
)}\,,
\end{split}
\ee
where $r_0$ is a theory-dependent rational number, $B$ (resp.\ $F$) denotes the bosonic (resp.\ fermionic) single-particle gauge-covariant letters, with the global charges $\{j_1,j_2\}$ (w.r.t.\ the Cartan generators of the $SO(2)_L\times SO(2)_R\subset SO(4)$ that parametrizes the angular momenta along $S^3$) and the collective gauge charges $\underline{\rho}\,$, and $p,q,\underline{z}$ are the corresponding fugacities.

\medskip

We consider general superconformal indices and will assume that the fugacities $(p,q)$ satisfy
\be\label{eq:ConditionPQOmega}
q^{\ell_q} \,=\, t^{{\ell_{t_q}}} \quad\textrm{and}\quad p^{\ell_p}\,=\, t^{{\ell_{t_p}}}\,, 
\qquad \textrm{with}\quad  \ell_{q},\ell_{p},\ell_{t_{p,q}}\in\,\mathbb{Z}_+\,,
\ee
in order to develop the Bethe ansatz formula below, and we wil call $t$ the \textit{multiplicative period}.

\smallskip

For the superconformal index, the integrand~$I$ can be shown to have an expansion around $t=0$ (and hence $q=0$ and $p=0$) of the form\footnote{
Again we have suppressed the dependence on other possible global charges.}
\be
\begin{split}
I\,=\,\sum_{m,n=0}^{\infty} I_{m,n}(\underline{z}) \,  q^{r m}\, p^{r n}   \qquad\textrm{with}\quad r\in \mathbb{Q}_+\,,
\end{split}
\ee
where the Taylor coefficients $I_{m,n}[\underline{z}]$ usually
have poles only at $\underline{z}=\underline{0}$ and~$\infty$. 
The rational number $r$ is related
to the rational number $r_0$ in \eqref{eq:IntegrandProd}, which can be understood as quantizing the eigenvalues of the charges $j_1$ and $j_2\,$; e.g.\ for 4D $\mathcal{N}=4$ SYM, which we will study later in the paper, $r=\frac{1}{3}\,$.

\bigskip

Consider the region $R$ defined by the following identifications of the gauge fugacities $z_i$:
\begin{equation}
z_{i}\,\sim\,\frac{z_i}{t^{n_i}}
\qquad \textrm{with} \quad 
|t|\,<\,1
\quad \textrm{and}\quad
n_i \,\in\,\mathbb{Z}\,,   
\end{equation}
in
\begin{equation}
(z_1,\ldots,z_{N-1}) \in \mathbb{C}^{N-1}\,.    
\end{equation}
Let~$\mathcal{C}_0$ be a mid-dimensional cycle of $R$ defined by the conditions
\begin{equation}\label{eq:C0Def}
|z_i|\,=\,1
\qquad\textrm{with}\quad i=1,\ldots N-1\,.    
\end{equation}
We define the \emph{shift operators} $\widehat{\mathcal{Q}}_j$ that implement the coordinate rescaling
\begin{equation}\label{eq:ShiftOperators}
\widehat{\mathcal{Q}}_{j}\,\star\, z_{i} \,
:=\, \frac{z_i}{t^{n^{(j)}_i}}\,, \qquad  j=1\,\ldots\, N-1 \,,
\end{equation}
with integer $n^{(j)}_i$'s. 
We can package the set  $\{n^{(j)}_i\}$ into  $N-1$ $(N-1)$-dimensional vectors (which we call \textit{periodicity vectors}):
\begin{equation}\label{eq:nji}
\underline{n}^{(j)}\,:=\,\{n^{(j)}_{i}\}\,,\qquad i\,=\,1\,,\ldots\,,\,N-1\,,
\end{equation}
or a single $(N-1)\times (N-1)$ matrix (which we call \textit{periodicity matrix}):
\begin{equation}\label{eq:nmatrixDef}
\begin{split}
\underline{\underline{n}}\,:=\,\{\underline{n}^{(j)}\}_{i=1,\ldots, N-1}\,:=\,\{n^{(j)}_i\}_{i,j=1\,,\ldots N-1}.
\end{split}
\end{equation}

\bigskip

We define the set of \textit{quasi-periodicities} of the integrand~$I$ under the action of the shift operators~$\widehat{\mathcal{Q}_j}$ as follows:
\be\label{eq:DefQuasiP}
\mathcal{Q}_{j}[\underline{z}]\, = \frac{\widehat{\mathcal{Q}}_j\star[\underline{z}] I[\underline{z}]}{I[\underline{z}]}\,:=\,\frac{ I[\widehat{\mathcal{Q}}_j\star\underline{z}]}{I[\underline{z}]}\,.
\ee
If the function $I[\underline{z}]$ is meromorphic in $\underline{z}\,$, then so are the functions $\mathcal{Q}_j$.

\subsection{Bethe expansion and Bethe solutions, for special Bethe contour}
\label{ssec:nSpecial}

It is convenient to first explain our formalism for the special choice of  
\begin{equation}\label{eq:ndelta}
n^{(j)}_i\,=\,\delta_{i,j}\,.  
\end{equation}
We will show how it applies to more general choices of integer $\underline{\underline{n}}$  later in Section~\ref{sec:nGeneral}.

\subsubsection{Bethe contour}
\label{sssec:nSpecial}

\medskip

The actions of the shift operators $\widehat{\mathcal{Q}}_j$ on the coordinates $z_i$, defined in~\eqref{eq:ShiftOperators}, naturally induces the actions of $\widehat{\mathcal{Q}}_j$ on the reference cycle $\mathcal{C}_0\,$, defined in \eqref{eq:C0Def}. 
For example, given the choice of integral vectors~\eqref{eq:ndelta}, this action produces new cycles
\begin{equation}
\mathcal{C}_{j} \,:=\, \widehat{\mathcal{Q}}_j\,\star\, \mathcal{C}_0\,:=\,\Biggl\{
\underline{z} \,\Bigl| \,
|z_{i}|=1 \textrm{ for }i\neq j \text{ and } |z_{j}|\,=\,\frac{1}{t}\,  \Biggr\}\,.    
\end{equation}
Furthermore, the action $\star$ also induces the product 
$\boldsymbol{\ast}$ between these cycles:
\begin{equation}
\begin{split}
\mathcal{C}_{j}
\,\boldsymbol{\ast}\,
\mathcal{C}_k\,& :=\, \widehat{\mathcal{Q}}_{j}\,\star\,\widehat{\mathcal{Q}}_{k}\, \star\,\mathcal{C}_0\,, 
\end{split}
\end{equation}
which gives 
\begin{equation}
\mathcal{C}_0
\,\boldsymbol{\ast}\,
\mathcal{C}_0\,=\, \mathcal{C}_0\,,    
\end{equation}
and products of multiple cycles by iteration.
The operation $\star$ is associative and distributive when acting over linear combinations of contours.
This allows us to define the \textit{fundamental annulus} $\mathcal{A}=\mathcal{A}(t)$ as follows
\be\label{eq:FundamentalAnnulus}
\mathcal{C}\,=\,\partial\mathcal{A}:=(\mathcal{C}_1 - \mathcal{C}_0) 
\,\boldsymbol{\ast}\, 
(\mathcal{C}_2 - \mathcal{C}_0)
\,\boldsymbol{\ast}\,
\ldots
\,\boldsymbol{\ast}\,
(\mathcal{C}_{N-1} - \mathcal{C}_0) \,,
\ee
which will play an  important role in the analysis below. 

Our formalism applies to theories whose meromorphic integrands~$I$ have the following properties.
\begin{property}\label{property1}
The quasi-periodicities~$\mathcal{Q}_{i}$ are invariant under the action of the shift operators:
\be\label{eq:PeriodicityQOper}
\widehat{\mathcal{Q}}_j\star \mathcal{Q}_{i}[\underline{z}]\,:=\, \mathcal{Q}_{i}[\widehat{\mathcal{Q}}_j\star\underline{z}]\,=\,\mathcal{Q}_i[\underline{z}]\,,\qquad \forall (i,j).
\ee
In particular, for superconformal theories, which are the focus of this paper, this condition can be satisfied using relations of the type~\eqref{eq:ConditionPQOmega}, see the review later. 
\end{property}

\begin{property}\label{property2}
Within the fundamental annulus $\mathcal{A}\,$, every $k$-multivariate pole $\underline{z}^{*}$ of $I(\underline{z})$  ($1\leq k<N$) is also a $k$-multivariate pole of the same order of the product $\prod_i \mathcal{Q}_i$, namely:
\be\label{eq:AssumptionPoles}
\frac{I(\underline{z})}{\prod_i \mathcal{Q}_i(\underline{z})} \text{  is regular at any such $\underline{z}^{*}$}\,.
\ee
For superconformal indices and for particular choice of $\mathcal{Q}_{i}$'s, the condition~\eqref{eq:AssumptionPoles} can always be achieved within appropriate domains of the flavour rapidities~\cite{Benini:2018mlo}. Using our result in Section.~\ref{sec:nGeneral}, it follows that their conclusion applies to any  possible choices of $\mathcal{Q}_i$ in such cases.  
\end{property}

\begin{property}\label{property3}
The condition~\eqref{eq:AssumptionPoles} together with definition~\eqref{eq:DefQuasiP} and the shift-symmetry of the $\mathcal{Q}_{i}$'s imply that any poles of~$I(\underline{z})$ outside the fundamental annulus $\mathcal{A}$ are generated by multiplication and division by prefactors $\mathcal{Q}_i$'s. 
Thus it follows that the poles of $I$ outside $\mathcal{A}$ are $t$-translated images of either zeroes or poles of $\mathcal{Q}_i\,$'s in~$\mathcal{A}$. 
\end{property}
Two remarks are in order here:
\begin{enumerate}
\item These three properties are satisfied by general superconformal theories upon imposition of conditions~\eqref{eq:ConditionPQOmega}. (It would be interesting to apply our method on systems other than superconformal indices that also  satisfy these two properties.)
\item These three properties hold for general integer period matrices $\underline{\underline{n}}$. 
\end{enumerate}

\subsubsection{Bethe expansion of indices}

Now we can define the Bethe expansion of the index, starting from its integral form~\eqref{eq:TheIntegral}.
Inserting the identity
\be
\frac{\prod_{i=1}^{N-1} (1 - \mathcal{Q}_i)}{\prod_{i=1}^{N-1} (1 - \mathcal{Q}_i)}\,=\,1\,,
\ee
into the integrand $I$,
expanding the numerator, using the properties of the action $\widehat{Q}_j$ on the right-hand side of~\eqref{eq:TheIntegral}, and the $t$-periodicity condition~\eqref{eq:PeriodicityQOper}, one reaches the following identity
\be\label{eq:BetheIdentity}
\begin{split}
&\oint_{\mathcal{C}_0} \prod_{i}\frac{\text{d}z_i}{2\pi\text{i}z_i}\, \cdot\,{I}[\underline{z}]
=\oint_{\mathcal{C}_0}\prod_{i}\frac{\text{d}z_i}{2\pi\text{i}z_i}\,\cdot\, \frac{\prod_i (1 - \mathcal{Q}_i)}{\prod_i (1 - \mathcal{Q}_i)}\, {I}[\underline{z}]
\\
&\qquad = \oint_{\mathcal{C}_0} \prod_{i}\frac{\text{d}z_i}{2\pi\text{i}z_i}\,\cdot\, \frac{1 - \mathcal{Q}_1 - \mathcal{Q}_2 +\ldots\,+\,(-1)^{N-1}\prod_i \mathcal{Q}_i}{\prod_i (1 - \mathcal{Q}_i)} {I}[\underline{z}]
\\ 
&\qquad = \Bigl( \oint_{\mathcal{C}_0} - \oint_{\mathcal{C}_1} - \oint_{\mathcal{C}_2} -\ldots
+(-1)^{N-1} \oint_{\mathcal{C}_1 \,\boldsymbol{\ast}\, 
\ldots
\,\boldsymbol{\ast}\, 
\mathcal{C}_{N-1}} \Bigr) \prod_{i}\frac{\text{d}z_i}{2\pi\text{i}z_i}\,\cdot\,\frac{{I}[\underline{z}]}{\prod_i (1 - \mathcal{Q}_i)}\,
\\ 
&\qquad = \oint_{(\mathcal{C}_1\,-\,\mathcal{C}_0) \,\boldsymbol{\ast}\,
\ldots
\,\boldsymbol{\ast}\, 
( \mathcal{C}_{N-1}-\mathcal{C}_0)} \prod_{i}\frac{\text{d}z_i}{2\pi\text{i}z_i}\,\cdot\, \frac{{I}[\underline{z}]}{\prod_i (\mathcal{Q}_i-1)}\,,
\end{split}
\ee
which gives the following: 

\begin{theorem}
Residue expansion
\be\label{eq:ResFormula}
\mathcal{I}\,=\,\sum_{\underline{z}^{*}\,\in\,\mathcal{A}} \text{Res}_{\mathfrak{m}_{\underline{z}^{*}}}\Biggl[\frac{{I}[\underline{z}]}{\prod_i z_i( \mathcal{Q}_i-1)}\,;\, \underline{z}^{*}\Biggr]\,,
\ee
where $\mathcal{A}=\mathcal{A}(t)$ is the fundamental annulus defined in ~\eqref{eq:FundamentalAnnulus}.

At last, from property~\ref{property2}, it follows that all the poles of the Bethe integrand that contribute to the residue formula~\eqref{eq:ResFormula} are solutions to the \textit{Bethe ansatz equations} (or simply \textit{Bethe equations}):
\be\label{eq:BAGeneral}
\mathcal{Q}_i\,=\,1\,,\qquad i\,=\,1\,,\ldots\,,\,
N-1\,.
\ee
The solutions $\underline{z}^{*}$ 
are called \textit{Bethe solutions} or \textit{Bethe roots}. Formula~\eqref{eq:ResFormula} is an example of \emph{Bethe expansion} of the superconformal index~$\mathcal{I}\,$.
\end{theorem}
\medskip
The integrand in the final form~\eqref{eq:BetheIdentity} will be called \emph{Bethe integrand}:
\be
I_{\textrm{B.A.}}=I_{\textrm{B.A.}}[\underline{\mathcal{Q}}]:=\frac{{I}[\underline{z}]}{\prod_i z_i( \mathcal{Q}_i-1)}\,.
\ee

The label $\mathfrak{m}_{\underline{z}^{*}}$ counts the number of possible values that the multivariate residue can take at~$\underline{z}^{*}\,$.\footnote{
For an introduction to the contour-selection dependence of multivariate residues in a related context the reader may refer to~\cite{Lee:2022vig}. }  (We will illustrate this ambiguity with an example in appendix~\ref{app:Ambiguity}.)
Note that this ambiguity is only for the individual residues; and the full index $\mathcal{I}$ (the left-hand side in residue formula~\eqref{eq:ResFormula}) is unambiguous. 
This is due to the fact that not all~$\mathfrak{m}_{\underline{z}^{*}}$ are independent, and the relations among them guarantee the final answer to be unambiguous.

The set of relations among~$\mathfrak{m}_{\underline{z}^{*}}$ are fixed by the shape of the Bethe contour $\mathcal{C}$ and the singularity structure of the integrand $I$. 
Determining these relations from such data, however, does not seem to be an easy geometric exercise. 
Instead of trying to do so geometrically, our strategy will be to leave the $\mathfrak{m}_{z^{*}}$'s unfixed and then determine the relations among them by first obtaining $\mathcal{I}\,$ using its alternative formula~\eqref{eq:TheIntegral}, and then imposing the identity~\eqref{eq:ResFormula}. 

\subsubsection{Isolated and continuous Bethe solutions}

If $\underline{z}^{*}$ is an \textit{isolated solution} to the Bethe equations, then there always exists a choice of $\mathfrak{m}_{\underline{z}^{*}}$ such that
\be\label{eq:BetheChoiceBM}
\text{Res}\Biggl[\frac{{I}[\underline{z}]}{\prod_i z_i(\mathcal{Q}_i-1)}\,;\, \underline{z}^{*}\Biggr]\,=\,\frac{{I}[\underline{z}^{*}]}{H[\underline{z}^{*}]}\,,
\ee
where $H[\underline{z}^{*}]$ is  the \emph{Jacobian} of a Bethe solution $\underline{z}^{*}$, defined as 
\be\label{eq:Jacobian}
\begin{split}
H[\underline{z}^{*}]&\,:=\,\text{det}_{i,j}\Biggl({z_j} \,\frac{\partial \mathcal{Q}_{ i}}{\partial{z_{ j}}} \Biggr)\Bigg|_{\underline{z}=\underline{z}^{*}}
\qquad \textrm{with}\quad i,j=1\,,\,\ldots\,,\,N-1\,.
\end{split}
\ee
For an isolated Bethe solution $\underline{z}^{*}$,
\begin{equation}
\textrm{isolated solution:}\qquad H[\underline{z}^{*}] \neq 0\,.    
\end{equation}
(Isolated solutions are also sometimes called non-degenerate solutions.)
Note that the choice of $\mathfrak{m}_{\underline{z}^*}$ in~\eqref{eq:BetheChoiceBM} was the assumption implicitly made in~\cite{Benini:2018mlo}.

\medskip

The Bethe equations~\eqref{eq:BAGeneral} can also have \textit{continuous} (i.e. non-isolated) Bethe solutions, which have the property:
\begin{equation}\label{eq:ContDef}
\textrm{continuous solution:}\qquad
H[\underline{z}^{*}] = 0   \,.
\end{equation}
(Continuous solutions are also sometimes called degenerate solutions.)
Because of the property~\eqref{eq:ContDef}, the equality~\eqref{eq:BetheChoiceBM} with definition~\eqref{eq:Jacobian} that apply to the isolated solutions do not apply to the continuous ones. 
One of the goals of this paper is to understand how the continuous Bethe solutions contribute to the Bethe expansion and thus to the index.

\subsection{General Bethe contours}\label{sec:nGeneral}

In the previous subsection, we have focused on the case where the periodicity matrix $\underline{\underline{n}}$ takes the special choice \eqref{eq:ndelta}.
In this subsection, let us consider more general choices of $n^{(j)}_i$.

\medskip

First of all, the new cycle created by the actions of $\widehat{\mathcal{Q}}_j$ on the reference cycle $\mathcal{C}_0\,$ becomes 
\begin{equation}
\mathcal{C}^{\underline{n}^{(j)}}_{j} \,=\, \widehat{\mathcal{Q}}_j\,\star\, \mathcal{C}_0\,:=\,\Biggl\{z_i\Bigl| \quad |z_{j}|\,=\,\frac{1}{t^{n^{(j)}_i}}\,.  \Biggr\}\,,
\end{equation}
and the products of these new cycles are similarly defined as 
\be
\begin{split}
\mathcal{C}^{\underline{n}^{(j)}}_{j}\,\boldsymbol{\ast}\, \mathcal{C}^{\underline{n}^{(k)}}_k\,& :=\, \widehat{\mathcal{Q}}_{j}\,\star\,\widehat{\mathcal{Q}}_{k}\, \star\,\mathcal{C}_0\,,
\end{split}
\ee
and the products of multiple cycles are defined iteratively. 
Correspondingly, for general $\underline{\underline{n}}$, its Bethe contour and annulus $\mathcal{A}$ are defined as
\be
\mathcal{C}^{\underline{\underline{n}}}\,=\,\partial\mathcal{A}^{\underline{\underline{n}}}:=({\boldsymbol{\ast}})_{j=1}^{N-1}(\mathcal{C}^{\underline{n}^{(j)}}_j-\mathcal{C}_0)
\,.
\ee

\medskip

The quasi-periodicities generated by the action of the shift operators $\widehat{\mathcal{Q}}_j$ on $I$ are
\be
\mathcal{Q}^{{\underline{\underline{n}}}}_j\,:=\,\prod_{i} (\mathcal{Q}_i)^{n^{(j)}_i}\,.
\ee
From properties~\ref{property2} and~\ref{property3} it follows that any multivariate pole $\underline{z}^{*}$ of $I$ within $\mathcal{A}^{\underline{\underline{n}}}$ is also a pole of the same order of $\prod_{i}\mathcal{Q}^{\underline{\underline{n}}}_i$, namely that
\be\label{eq:AssumptionPoles2}
\frac{I(\underline{z})}{\prod_i \mathcal{Q}^{\underline{\underline{n}}}_i(\underline{z})} \text{  is regular at any such $\underline{z}^{*}$}\,.
\ee
Consequently, for each allowed choice of $\underline{\underline{n}}$, we obtain:

\begin{theorem}
Generalized Bethe expansions
\be
\mathcal{I}\,=\,\int_{\mathcal{C}^{\underline{\underline{n}}}} \prod_i\frac{\text{d}z_i}{2\pi\text{i}z_i} \cdot \frac{I[\underline{z}]}{\prod_j(\mathcal{Q}^{\underline{\underline{n}}}_j-1)}\,=\,\sum_{\underline{z}^{*}\,\in\,\mathcal{A}^{\underline{\underline{n}}}} \text{Res}_{\mathfrak{m}_{\underline{z}^{*}}}\Biggl[\frac{{I}[\underline{z}]}{\prod_i z_i( \mathcal{Q}^{\underline{\underline{n}}}_i-1)}\,;\, \underline{z}^{*}\Biggr]\,,
\ee
where the $\underline{z}^{*}$ are solutions to a generalized set of Bethe equations
\be
\mathcal{Q}^{\underline{\underline{n}}}_j\,:=\,\prod_{i} (\mathcal{Q}_i)^{n^{(j)}_i}\,=\,1\,,\qquad j\,=\,1\,,\ldots \,N-1\,.
\ee
\end{theorem}

Note that every solution to $\mathcal{Q}_i=1$ is a solution to $\mathcal{Q}^{\underline{\underline{n}}}_j=1$, but not necessarily the other way around. 
For the total set of solutions to coincide, we need  that 
\be
|\det{\underline{\underline{n}}}|\,=\,1\, \quad\equiv\quad\underline{\underline{n}} \,\in\,{GL}(N-1,\mathbb{Z})\,.
\ee
It is however sufficient to consider those $\underline{\underline{n}}$ inside the signed unipotent subgroup of upper-triangular matrices $U^*(N-1,\mathbb{Z})\subset GL(N-1,\mathbb{Z})$:
\be\label{eq:SignedUnipotent}
{
\underline{\underline{n}} = \begin{pmatrix}
\pm 1 & a_{1,1} & a_{1,2} & \cdots & a_{1,N-2} \\
0 & \pm 1 & a_{2,2} & \cdots & a_{2,N-2} \\
0 & 0 & \pm 1 & \cdots & a_{3,N-2} \\
\vdots & \vdots & \vdots & \ddots & \vdots \\
0 & 0 & 0 & \cdots & \pm 1
\end{pmatrix}\,,\qquad a_{i,j}\,\in\,\mathbb{Z}\,.}
\ee

\subsubsection{Ambiguity in the choice of Bethe integrand}\label{sec:amb211}
The choice of Bethe integrand $I_{\textrm{B.A.}}[\underline{\mathcal{Q}}]$ is ambiguous. 
For a given Bethe contour $\mathcal{C}$, there still exist different choices of $\underline{\mathcal{Q}}$ that keep the formula invariant. 
Let us proceed to show this.

For example, the following identity
\be\label{eq:QQMinusOneEquiv}
\begin{split}
\int_{\mathcal{C}^{\underline{\underline{n}}}} \prod_i\frac{\text{d}z_i}{2\pi\text{i}z_i} \cdot \frac{I[\underline{z}]}{\prod_j (\prod_{i} (\mathcal{Q}_i)^{n^{(j)}_i}-1)}&\,=\,\int_{(-1)^{N-1}\mathcal{C}^{\underline{\underline{n}}}} \prod_i\frac{\text{d}z_i}{2\pi\text{i}z_i} \cdot \frac{\prod_j\prod_{i} (\mathcal{Q}_i)^{n^{(j)}_i}\cdot I[\underline{z}]}{\prod_j (\prod_{i} (\mathcal{Q}_i)^{n^{(j)}_i}-1)} \\ &\,=\,\int_{(-1)^{N-1}\mathcal{C}^{\underline{\underline{n}}}} \prod_i\frac{\text{d}z_i}{2\pi\text{i}z_i} \cdot \frac{ I[\underline{z}]}{\prod_j (1-\prod_{i} (\mathcal{Q}_i)^{-n^{(j)}_i})} \\&\,=\,\int_{\mathcal{C}^{\underline{\underline{n}}}} \prod_i\frac{\text{d}z_i}{2\pi\text{i}z_i} \cdot \frac{ I[\underline{z}]}{\prod_j (\prod_{i} (\mathcal{Q}_i)^{-n^{(j)}_i}-1)}
\end{split}
\ee
where 
\be
(-1)^{N-1}\mathcal{C}^{\underline{\underline{n}}}\,:=\,({\boldsymbol{\ast}})_{j=1}^{N-1}(-\mathcal{C}^{\underline{n}^{(j)}}_j+\mathcal{C}_0)\,
\ee
implies that at fixed $\underline{\underline{n}}\,$, the Bethe expansion with~$\mathcal{Q}_j$ inserted is equivalent to the one with $\mathcal{Q}_j^{-1}$ inserted.

\begin{theorem}
More generally, the integrand of the Bethe formula
\be
\mathcal{I}=\int_{\mathcal{C}} \prod_i\frac{\text{d}z_i}{2\pi\text{i} z_i} \cdot \frac{I[\underline{z}]}{\prod_i(\mathcal{Q}_i[\underline{z}]-1)}\,,
\ee
deduced with $\underline{\underline{n}}=\{\delta_{ij}\}$ and $\mathcal{Q}_i \,$, can be deformed by substituting\footnote{
The open index labels the column of $\underline{\underline{n}}^\prime$.}
\be
\mathcal{Q}_i \,\to\, \mathcal{Q}^{\underline{\underline{n}}^\prime}_i=\prod_{j}(\mathcal{Q}_j)^{\underline{\underline{n}}^\prime_{j, i}}\,, 
\ee
where $\underline{\underline{n}}^\prime$ is a signed unipotent matrix of dimension $N-1\,$~\eqref{eq:SignedUnipotent}. Namely,
\be\label{eq:IntegrandDefStatement}
\mathcal{I}=\int_{\mathcal{C}} \prod_i {\text{d} z_i} \cdot I_{\textrm{B.A.}}[{\mathcal{Q}}]=\int_{\mathcal{C}} \prod_i {\text{d} z_i} \cdot I_{\textrm{B.A.}}[
{\mathcal{Q}^{\underline{\underline{n}}^\prime}}]\,.
\ee
\end{theorem}
To prove this, we implement a signed unipotent change of integration coordinates
\be\label{eq:UnipotentChangCoords}
\underline{z} \to \underline{w}=\underline{f}(\underline{z}) \qquad \text{s.t.} \qquad \underline{f}(\mathcal{C}^{\underline{\underline{n}}})\,=\,\mathcal{C} \text{  and  } \underline{f}(\mathcal{A}^{\underline{\underline{n}}})\,=\,\mathcal{A}\,.
\ee
Concretely
\begin{equation}
f^{-1}_i(\underline{w})\,:=\,\prod_j(w_j)^{\underline{\underline{n}}_{j,i}}\,=\,e^{\sum_{j}\underline{\underline{n}}_{j,i}\log {w}_j}\,.
\end{equation}
It will be useful to recall that
\be
\mathcal{Q}^{\underline{\underline{n}}}_i (\underline{z}) \,=\, \prod_{j}(\mathcal{Q}_j)^{\underline{\underline{n}}_{j,i}}= e^{\sum_{j}\underline{\underline{n}}_{j,i}\log \mathcal{Q}_j}\,.
\ee
Under this change of variables, the measure changes to
\be
\prod_i\frac{\text{d}z_i}{2\pi \text{i} z_i}\,=\,|\det{\underline{\underline{n}}}|\, \prod_i\frac{\text{d}w_i}{2 \pi \text{i}w_i}\,.
\ee
and
\be
\begin{split}
\int_{\mathcal{C}^{\underline{\underline{n}}}} \prod_i\frac{\text{d}z_i}{2\pi\text{i}z_i} \cdot \frac{I[\underline{z}]}{\prod_j(\mathcal{Q}_j^{\underline{\underline{n}}}-1)}&
\,=\,
|\det{\underline{\underline{n}}}|\int_{\mathcal{C}} \prod_i\frac{\text{d}w_i}{2\pi\text{i}w_i} \cdot \frac{I[\underline{f}^{(-1)}(\underline{w})]}{\prod_j(\mathcal{Q}_j^{\underline{\underline{n}}}[\underline{f}^{(-1)}(\underline{w})]-1)} \\
&\,=\,
\sum_{\underline{z}^{*}\,\in\,\mathcal{A}} \text{Res}_{\mathfrak{m}_{\underline{z}^{*}}}\Biggl[\frac{{I}[\underline{f}^{(-1)}(\underline{w})]}{\prod_i w_i( \mathcal{Q}^{\underline{\underline{n}}}_i[\underline{f}^{(-1)}(\underline{w})]-1)}\,;\, \underline{z}^{*}\Biggr]\,.
\end{split}
\ee
If one assumes only isolated poles then the series of relations follow
\be\label{eq:DetailsQuasiInvariant}
\begin{split}
\frac{\mathfrak{m}_{z^{*}}I[\underline{f}^{(-1)}(\underline{z}^{*})]}{\det{ \partial_{\log{w_i}}\log\mathcal{Q}^{\underline{\underline{n}}}_{j}}[\underline{f}^{(-1)}(\underline{z}^{*})]]
}&\,=\,\frac{\mathfrak{m}_{z^{*}} I[\underline{z}^{*}]}{\det{ \partial_{\log{w_i}}\log\mathcal{Q}^{\underline{\underline{n}}}_{j}}[\underline{f}^{(-1)}(\underline{z}^{*})]]
} \\
&\,=\,\frac{\mathfrak{m}_{z^{*}}I[\underline{z}^{*}]}{\det{ \partial_{\log{z_i}}\log\mathcal{Q}^{\underline{\underline{n}}}_{j}}[\underline{z}^{*}]
}\, \frac{1}{\det{\underline{\underline{n}}}}\\
&\,=\,\frac{\mathfrak{m}_{z^{*}}I[\underline{z}^{*}]}{\det{ \partial_{\log{z_i}}\log\mathcal{Q}_{j}}[\underline{z}^{*}]
}\, \frac{1}{(\det{\underline{\underline{n}}})^2}\\ 
&\,=\,\frac{\mathfrak{m}_{z^{*}} I[\underline{z}^{*}]}{\det{ \partial_{\log{z_i}}\log\mathcal{Q}_{j}}[\underline{z}^{*}]
}\, .
\end{split}
\ee
In these relations we have used that $f^{-1}(\underline{z}^{*})$ and $\underline{z}^{*}$ are identified upon rescaling $\underline{z}\sim \underline{z}t\,$, which are isometries of $\mathcal{Q}_i^{\underline{\underline{n}}}$ at any $\underline{z}$, and of $I$ at the locus $\mathcal{Q}^{\underline{\underline{n}}}_i=1\,$, respectively.

As $\underline{\underline{n}}$ belongs to~\eqref{eq:SignedUnipotent}, the sets of $\underline{z}^{*}$'s solving
\be
\mathcal{Q}^{\underline{\underline{n}}}_i=1\qquad  \text{   and   } \qquad \mathcal{Q}_i=1 
\ee
are equivalent. 
Summing over the solutions $\underline{z}^{*}$, one concludes~\eqref{eq:IntegrandDefStatement}
\be\label{eq:ContourChange}
\mathcal{I}\,=\,\int_{\mathcal{C}} \prod_i\frac{\text{d}z_i}{2\pi\text{i}z_i} \cdot \frac{I[\underline{z}]}{\prod_j(\mathcal{Q}_j[\underline{z}]-1)}\,=\,\int_\mathcal{C} \,\prod_{i}\,\frac{\text{d}z_i}{2\pi\text{i}z_i} \,\cdot\,\frac{I[\underline{z}]}{\prod_j \bigl(\mathcal{Q}^{\underline{\underline{n}}}_j[\underline{z}]-1\bigr)}\,.
\ee
Note that this analysis relies on the assumption of the absence of continuous solutions. 
Next, we will show that the conclusion remains the same even in the presence of the continuous solutions.

\subsection{Contribution from continuous solutions} \label{subsec:MethodContinuous}

The continuous Bethe solutions can only exist for $N\,\geq\,3\,$ \cite{Lezcano:2021qbj,Benini:2021ano}. 
For simplicity we will focus on the case $N=3\,$. The discussion for larger values of $N$ is more involved, but the idea is essentially the same. 

\medskip

For $N=3$ there are two independent complex variables within the vector~$\underline{z}\,$, therefore in this case, the continuous solutions, if exist, can only have complex dimension $1\,$. 
Namely, they can be defined by a complex curve
\be
\gamma\,:=\,\bigl\{\underline{z}\,\in\,\mathbb{C}^2\,:\,\underline{z}=\underline{z}(s)\bigr\}\,,
\ee
with the world-sheet parameter~$s\in \mathbb{C}\,$.
Concretely, the curve $\gamma$ can be defined by the vanishing of a meromorphic map~$h:\mathbb{C}^2\mapsto \mathbb{C}$
\be
 h(\underline{z}=\underline{z}(s))\,=\,0\,.
\ee
Let us now examine the condition under which the complex solution exist and how to define $h$ from the Bethe operators $\mathcal{Q}_i$, $i=1,2$.

\medskip

Consider the factorization
\begin{equation}\label{eq:Defh}
(1\,-\,\mathcal{Q}_i(\underline{z}))\,=:\,h(\underline{z})  \, \mathcal{Q}^{(1)}_i(\underline{z})\qquad \textrm{with}\quad
i\,=\,1\,,\,2\,.
\end{equation}
\begin{enumerate}
\item A continuous solution (with $\textrm{dim}_{\mathbb{C}}=1$) exist if $h$ is a ``common divisor" of the two denominator factors, $( 1-\mathcal{Q}_i)$, $i=1,2$.
Namely, $h$ is a meromorphic map: $\mathbb{C}^2\mapsto \mathbb{C}$, which defines the complex curve $\gamma$, and the two meromorphic maps $\mathcal{Q}_{i}^{(1)}\,:\mathbb{C}^2\mapsto \mathbb{C}$, defined as the quotients
\be
\frac{(1\,-\,\mathcal{Q}_i(\underline{z}))}{h(\underline{z})}\,,
\ee
is \textit{finite} on $\gamma$.

\item We also require that $h$ is the ``largest" common divisor of the two denominator factors, $( 1-\mathcal{Q}_i)$, $i=1,2$. 
(This is to include all continuous factors and to uniquely fix $h$.)
Namely, in addition to the condition earlier, $\mathcal{Q}_{i}^{(1)}$ must also be \textit{non-vanishing} on $\gamma$ except for at two \textit{distinct} sets of \textit{isolated} intersecting points
\be
P_{i}\,=\,\{\underline{\mathfrak{z}}^{\ast(i)}_1\,,\underline{\mathfrak{z}}^{\ast(i)}_2\,,\,\ldots\}\,\,\qquad i\,=\,1\,,\,2\,,
\ee
defined by the intersection conditions
\be
h(\underline{\mathfrak{z}}^{\ast (i)})
\,=\,\mathcal{Q}_{i}^{(1)}(\underline{\mathfrak{z}}^{\ast (i)})\,=\,0\,, \qquad  i\,=\,1,\,2\,,
\ee
together with the null intersection condition:
\be\label{eq:NullIntersection}
P_{1}\, \cap\, P_{2} \,=\, 0\,.
\ee
Note that the null-intersection condition~\eqref{eq:NullIntersection} is always achievable with the proper definition of $h(\underline{z})\,$. 
Namely, if one starts from a definition of $h$ for which there exists a common solution to $h=Q_1^{(1)}=Q_2^{(1)}=0$, then there must exist a common divisor $h_2=h_2(\underline{z})$ to both $\mathcal{Q}^{(1)}_i(\underline{z})\,$ that is not a divisor of $h\,$. 
In this case, one can always redefine 
\be
h\,\to\, h h_2 \,\implies\, \mathcal{Q}^{(1)}_i \,\to\, \frac{\mathcal{Q}^{(1)}_i}{h_2}\,.
\ee
With these redefinitions~\eqref{eq:NullIntersection} is achieved.
\end{enumerate}

\smallskip

To summarize, for the case of $N=3$, the $h$ and $\mathcal{Q}_i$ that satisfy the properties above classify the Bethe roots $\underline{z}^{\ast (i)}$ into:
\begin{enumerate}
\item Set of isolated solutions $\underline{z}=\underline{\boldsymbol{z}}^{\ast}$:
\begin{equation}
\mathcal{Q}_{1}^{(1)}(\underline{\boldsymbol{z}}^{\ast (i)})
\,=\,\mathcal{Q}_{2}^{(1)}(\underline{\boldsymbol{z}}^{\ast (i)})
\,=\, 0\,.    
\end{equation}
\item $\gamma$: set of continuous solutions $\underline{z}=\underline{\mathbf{z}}^{\ast}$
\begin{equation}
\begin{aligned}
&h(\underline{\mathbf{z}}^{\ast})=0\\
\textrm{and} \quad &
\mathcal{Q}_{1,2}^{(1)}(\underline{\mathbf{z}}^{\ast})
\,\neq\, 0 \quad \textrm{except for isolated points } \underline{\mathbf{z}}^{\ast}=\underline{\mathfrak{z}}^{\ast}       
\end{aligned}
\end{equation}
\begin{enumerate}
\item $P_1\cup P_2:$ isolated points $\underline{z}=\underline{\mathfrak{z}}^{\ast}$ on $\gamma$ satisfying
\begin{equation}
h(\underline{\mathfrak{z}}^{\ast (i)})
\,=\,\mathcal{Q}_{i}^{(1)}(\underline{\mathfrak{z}}^{\ast (i)})\,=\,0\,, \qquad  i\,=\,1  \textrm{ or } 2\,,
\end{equation}
\item $\gamma\backslash P_1\cup P_2:$ generic points $\underline{\mathbf{z}}^{\ast}$ on $\gamma\,$.
\end{enumerate}
\end{enumerate}

\bigskip
\begin{theorem}
From the continuous solution $\gamma$, only the discrete set of isolated points $P_{1}\cup P_{2}$ contributes to the integral~$\mathcal{I}\,$. 
\end{theorem}

Let us now show this. 
We can use the identity
\be
 \begin{split}
 \frac{I}{(1 - \mathcal{Q}_1)(1 - \mathcal{Q}_2)} \frac{dz_1}{z_1} \wedge \frac{dz_2}{z_2}  &= \frac{I}{( \mathcal{Q}_1-1)h \mathcal{Q}^{(1)}_2} \,\frac{dz_1}{z_1} \wedge \frac{dz_2}{z_2} 
 \end{split}\,.
\ee
We then introduce the following change of coordinates in an infinitesimal vicinity of $\gamma$:
\be
\underline{z} \,\mapsto\, \Bigl({\mathcal{Q}}_1\,, \mathcal{Q}_{2}^{(1)}\Bigr)\,.
\ee
In these local coordinates, and after a Taylor expansion in the numerator, the integrand 2-form reduces to
\be
\ldots\,+\,\frac{1}{( \mathcal{Q}_1-1) \mathcal{Q}^{(1)}_2} \partial_h\left(\frac{{{I}}}{H_{1,2}}\, \right) d\mathcal{Q}_1 \wedge d \mathcal{Q}^{(1)}_2\,+\,\dots\,,
\ee
where~
\be
\left( \partial_{h} f \right)\,:=\,\sum_i \frac{\partial z_i}{\partial h(\underline{z})} \partial_{z_i} f[\underline{z}]
\ee
denotes the gradient of $f$ normal to the regions
\be
h(\underline{z})\,=\, h\,=\,\text{const}\,,
\ee
and the Jacobian $H_{1,2}$ is defined by specializing $i=1$ and $j=2$ in
\be
 H_{i,j}\,:=\,\frac{\partial (\log \mathcal{Q}_i,\mathcal{Q}^{(1)}_j)}{\partial (\log z_1, \log z_2)}\,\neq\,0\,. 
 \ee
These local coordinates can be used to compute the residues of all the isolated points $\underline{\mathfrak{z}}^{\ast}\in P_2\,$. The answer is
\be\label{eq:resP2}
\text{Res}_{\mathfrak{m}_{\underline{\mathfrak{z}}^{*}}}[I_{\textrm{B.A.}};\underline{\mathfrak{z}}^{*} \in P_2]\,:=\,{\mathfrak{m}_{\underline{\mathfrak{z}}^{*}}}\partial_h\left(\frac{{{I}}}{H_{1,2}}\, \right)\Biggl|_{h=\mathcal{Q}^{(1)}_2=0}\,.
\ee
Analogously, for the residues of all the isolated points $\underline{\mathfrak{z}}^{\ast}\in P_1\,$, the answer is
\be\label{eq:resP1}
\text{Res}_{\mathfrak{m}_{\underline{\mathfrak{z}}^{*}}}[I_{\textrm{B.A.}};\underline{\mathfrak{z}}^{*}\,\in\,P_1]\,:=\, {\mathfrak{m}_{\underline{\mathfrak{z}}^{*}}} \partial_h\left(\frac{{{I}}}{H_{2,1}}\, \right) \Biggl|_{h=\mathcal{Q}^{(1)}_1=0}\,.
\ee

In section~\ref{sec:5} we will study a concrete example where there are~$5$ different possible $\gamma$'s and~$h(z)$'s. 
This will be reported in table~\ref{tab:two_columns}. 
In that example, the~$h(\underline{z})$'s, and their corresponding $\mathcal{Q}^{(1)}_{i}$'s, will be computed in asymptotic expansions around $t=0\,$. 
The application of the residue formulae~\eqref{eq:resP2} and~\eqref{eq:resP1} in these cases reproduce the results independently derived in section~\ref{sec:5}.

\bigskip

Finally, we note that under the signed unipotent transformations of integration variables $\underline{z}$ defined in equation~\eqref{eq:UnipotentChangCoords}, the quantities
\be
\partial_h\left(\frac{{{I}}}{H_{i,j}}\right)
\ee
transform as a quasi-scalar
\be
\partial_h\left(\frac{{{I}}}{H_{i,j}}\right) \,\longrightarrow \,\frac{1}{\det \underline{\underline{n}}}\partial_h\left(\frac{{{I}}}{H_{i,j}}\right)\,,
\ee
(the independent variable within each function involved, non-trivially transformed). 
This follows from analogous steps to the ones detailed in equation~\eqref{eq:DetailsQuasiInvariant}. 
Then restricting the internal independent variables to the isolated Bethe solution $\underline{z}=\underline{\mathfrak{z}}^{*}$
\be
h(\underline{\mathfrak{z}}^{*})=\mathcal{Q}^{(1)}_{j}(\underline{\mathfrak{z}}^{*})=0 \implies (\mathcal{Q}_{1,2
}(\underline{\mathfrak{z}}^{*})-1)\,=\,0
\ee
and using the fact that $h$ and $\mathcal{Q}^{(1)}_i$ functions\footnote{
We are assuming here the global versions of them, not just some truncation of their asymptotic expansions around $t=0$, for instance.} are periodic under $t$-shifts and $I$ is periodic under $t$-shifts on Bethe solutions, we conclude that under the transformations~\eqref{eq:UnipotentChangCoords}, the residues
\be
{\mathfrak{m}_{\underline{\mathfrak{z}}^{*}}}\partial_h\left(\frac{{{I}}}{H_{i,j}}\, \right)\Biggl|_{h=\mathcal{Q}^{(1)}_j=0}
\ee
are quasi-invariants
\be\label{eq:QUasiInv}
{\mathfrak{m}_{\underline{\mathfrak{z}}^{*}}}\partial_h\left(\frac{{{I}}}{H_{i,j}}\right)\Biggl|_{h=\mathcal{Q}^{(1)}_j=0} \longrightarrow \frac{{\mathfrak{m}_{\underline{\mathfrak{z}}^{*}}}}{\det \underline{\underline{n}}}\,\partial_h\left(\frac{{{I}}}{H_{i,j}}\right)\Biggl|_{h=\mathcal{Q}^{(1)}_j=0}\,,
\ee
precisely as the residues of isolated solutions (as in the second line of equation~\eqref{eq:DetailsQuasiInvariant}).

\bigskip

A signed unipotent transformation
of the Bethe operators
\be\label{eq:ReplacementCont}
{\mathcal{Q}}_i \,\to\, {\mathcal{Q}}^{\underline{\underline{n}}}_i,
\ee
induces the transformation:
\be\label{eq:InducedTransf}
\begin{split}
I\to I\,,\qquad h\to h\,,\qquad H_{i,j}\to \det \underline{\underline{n}} \,H_{i,j}\,,
\end{split}
\ee
without transforming the integration variables. Let us explain. 

First, $I$ does not depend on the choice of Bethe operator $\mathcal{Q}_i$'s, therefore it remains invariant under the replacements~\eqref{eq:ReplacementCont}.

Second, the set of Bethe solutions $\underline{z}^{\ast}$ remains invariant under the replacements~\eqref{eq:ReplacementCont}, including continuous solutions $\underline{\mathfrak{z}}^\ast$.
Consequently, the common divisor $h=h(\underline{z})$ also remains invariant under the replacements~\eqref{eq:ReplacementCont}. 

Third, the transformation of the determinants $H_{i,j}$ follows from the transformations
\be
\begin{split}
\{\log \mathcal{Q}_1, \frac{\log\mathcal{Q}_2}{h}\}\,=\,\{\log \mathcal{Q}_1, \mathcal{Q}_2^{(1)}\}\to \{\log \mathcal{Q}_1, \mathcal{Q}_2^{(1)}\}\cdot \underline{\underline{n}}\,,\\
\{\frac{\log\mathcal{Q}_1}{h},\log \mathcal{Q}_2\}\,=\,\{\mathcal{Q}_1^{(1)},\log \mathcal{Q}_2\}\to \{ \mathcal{Q}_1^{(1)},\log \mathcal{Q}_2\}\cdot \underline{\underline{n}}\,,
\end{split}
\ee
where the equalities in the first and second lines hold in the infinitesimal vicinities of~$\underline{\mathfrak{z}}^\ast$'s in $P_2$ and $P_1$, respectively.

Finally, as $(\det \underline{\underline{n}})^2=1$, the composition of~\eqref{eq:QUasiInv} and~\eqref{eq:InducedTransf} completes the proof of~\eqref{eq:ContourChange} (for $N=3$).

\medskip

For $N\,>\,3$, the analysis becomes more involved because the continuous solutions might be manifolds with complex dimension larger than $1\,$. 
In such cases there might be many more variants, and the number of them increases with $N\,$. 
Those more general cases will be analyzed elsewhere. 
The idea to approach them is essentially the same as the one just explained. 
In particular, by the obvious generalization of the localization mechanism just reported, all such continuous solutions are bound to localize to non-degenerate roots (if they contribute at all).

\subsection{A method to find Bethe solutions}\label{eq:RecursiveMethod}

In an expansion at small rapidity $t$, one may study how the Bethe discontinuity $\mathcal{Q}_i$ behaves using the $t$-independent Taylor coefficient functions $I_{m,n}$
\be
\begin{split}
\mathcal{Q}_i(\underline{z}) \,=\, \frac{\sum_{m,n\,=\,0}^{\infty}{I}_{m,n}(\ldots,\frac{{z}_i}{t}\,,\ldots)\,q^{mr} p^{n r} }{\sum_{m,n\,=\,0}^{\infty}{I}_{m,n}(\underline{z})\,q^{mr} p^{n r}}\,.
\end{split}
\ee
Here the numerator
\be
{\sum_{m,n\,=\,0}^{\infty}{I}_{m,n}(\ldots,\frac{{z}_i}{t}\,,\ldots)\,q^{mr} p^{n r}}
\ee
should be viewed as a formal power series.\footnote{
For computations we truncate this expansion at a convenient order.}
  
The following property
\be\label{eq:LargeZ}
\forall_{j\,\neq\, i=1\,,\ldots,\,N-1} I(\underline{z}) \,\underset{z_{i}\to\infty\atop z_{j}, q, p \text{ fixed }}{=}\,\mathcal{O}(z_{i}^0)\,,
\ee
implies the cancellation of all singularities coming from expanding the individual Taylor coefficient functions~$I_{m,n}(\ldots\,,\,\frac{{z}_i}{t}\,,\ldots)$ around~$t=0\,$ at fixed $z_{j}\,$'s for all $j\,\neq\, i\,$.
The property~\eqref{eq:LargeZ} holds for theories with the same number of bosonic and fermionic single-particle states at each gauge charge vector $\underline{\rho}\,$, which is the case of superconformal theories. From~\eqref{eq:LargeZ} and the same reason it follows that
\be\label{eq:BalancingDelta0}
\mathcal{Q}_{i}(\underline{z})\,:=\,\frac{\sum_{m,n\,=\,0}^{\infty}{I}_{m,n}(\ldots,\frac{{z}_i}{t}\,,\ldots)\,q^{mr} p^{n r} }{\sum_{m,n\,=\,0}^{\infty}{I}_{m,n}(\underline{z})\,q^{mr} p^{n r}}\,\underset{t\,\to\,0\atop\underline{z} \text{ fixed}}{=}\,\mathcal{O}(t^{0})\,.
\ee

We call \emph{primary spectrum} of Bethe solutions the set of all possible $(N-1)$- real vectors
\be
\underline{\Delta}=\Bigl\{\Delta_i\,\Bigl|0\,\leq\,\text{sign}(n^{(j)}_i)\,\Delta_i \,<\,|n^{(j)}_i|\Bigr\}_{i=1,\ldots N-1}\,,
\ee
such that, for
\be\label{eq:SmallQExpansion}
z_i \underset{t\,\to\,0}{=} Z_i t^{-\Delta_i}\,, \quad Z_i\,=\, Z_i^{(0)}\,+\,\ldots\, 
\ee
(with the~$Z^{(0)}_i$ being complex constants) one obtains
\be\label{eq:BalancingCondition}
\mathcal{Q}_i(\underline{z})\,\underset{t\,\to\, 0}{=}\, \mathcal{R}_{\underline{\Delta},i}[\underline{Z}^{(0)}]\,+\,\ldots\,,
\ee
with the condition
\be\label{eq:VanishingBetheTrig}
\mathcal{R}_{\underline{\Delta},i}[\underline{Z}^{(0)}]=0
\ee
having non-vanishing solutions $\underline{Z}^{(0)}\,$. 
The
$\ldots$ denote subleading corrections in the corresponding small-$t$ expansion.

Note that~\eqref{eq:BalancingCondition} does not follow from~\eqref{eq:BalancingDelta0} except when~$\underline{\Delta}=0\,$. 
Indeed, from~\eqref{eq:BalancingDelta0} it follows that for superconformal theories $\underline{\Delta}=0$ is always part of the spectrum of primary weights provided the corresponding condition~\eqref{eq:VanishingBetheTrig} has non-vanishing solutions.

Given any such $\underline{\Delta}$, the
\be
\mathcal{R}_{\underline{\Delta},i}\,:=\,\mathcal{R}_{\underline{\Delta},i}[\underline{Z}^{(0)}] 
\ee
are $N-1$ $t$-independent rational functions of $\underline{Z}^{(0)}\,$.

\begin{theorem}
The solutions of the original Bethe equations
\be\label{eq:BAEqParticular}
\mathcal{Q}_i\,=\,1
\ee
are in one-to-one relation with solutions of the rational equations
\be\label{eq:TrigoBetheEqs}
\mathcal{R}_{\underline{\Delta},i}\,=\,1\,, \qquad \forall \underline{\Delta}\,.
\ee
\end{theorem}
For the examples that we have studied in this paper which correspond to the particular cases $N=3,4,5$ (we have not proven this in general), for a given $\underline{\Delta}\,$, the corresponding rational system of $N-1$ equations for the $N-1$ variables $Z^{(0)}_i$ can be reduced to a finite set of independent order-$N$ polynomial equations in a single complex variable $Z$ 
\be
\mathcal{P}^{(0)}_{\underline{\Delta}}(Z)\,=\,0\,\quad \lor \quad \mathcal{P}^{(1)}_{\underline{\Delta}}(Z)\,=\,0\quad \lor \quad \ldots .
\ee
$Z$ can be thought of as one of the original $Z^{(0)}_i$, the remaining one after having solved $N-2$ out of the $N-1$ equations in~\eqref{eq:TrigoBetheEqs}.

Generically, the Galois group of the independent $\mathcal{P}^{(r)}_{\underline{\Delta}}$'s is the permutation group $S_{N}\,$. 
When the latter is broken, a discrete symmetry emerges among the polynomial roots. 
For example, for the solutions determined by the symmetry operations $z_i \sim z_i t$ and $\Xi_k$, a $\mathbb{Z}_{N}$ symmetry is there. Their defining polynomial takes the very simple form 
\be
\mathcal{P}^{(0)}_{\Delta}\,=\, 1-Z^{N}\,.
\ee
In the examples we have studied, solutions corresponding to other polynomials $\mathcal{P}^{(r)}_{\underline{\Delta}}$ happens to have Galois group $S_{N}$ at generic values of the flavour rapidities $Y_x\,$.

The complete form of each independent solution can be reconstructed by solving~\eqref{eq:BAEqParticular} in the small-$t$ expansion determined by a given primary~$\Delta$~\eqref{eq:SmallQExpansion} around the independent solutions fixed by the individual polynomials~$\mathcal{P}^{(r)}_\Delta\,$. However, for topological solutions the perturbative expansion truncates at leading order.
Non-topological solutions do not truncate, but it turns out that to compute residue contributions in the small-$t$ expansion one can effectively work with the very leading contribution to the position of the poles $Z_i^{(0)}$ -- the positions of the points where the contribution of the solution-manifold localizes at -- and forget about the subleading contributions.

\section{Our working example: $\mathcal{N}=4$ Super-Yang-Mills}\label{sec:3}

In this section, we apply the  formalism introduced in the previous section to study the superconformal index of the $\mathcal{N}=4$ SYM theory with ~$U(N)$ or $SU(N)$ gauge group. Throughout this paper, we adopt the notation
\begin{equation}
\textrm{e}(x)\,:=\, e^{2\pi \text{i} x}
\,.
\end{equation}

\subsection{Superconformal index of $\mathcal{N}=4$ SYM}

For the 4D $\mathcal{N}=4$ SYM, the superconformal index, defined as the Witten index of the theory in the radial quantization, is given by (following the conventions of~\cite{Benini:2018mlo}) 
\begin{equation}
\mathcal{I}\,:=\,
\mathrm{Tr}_{\mathcal{H}} \left[
(-1)^F 
e^{-\beta \lbrace \mathcal{Q},\mathcal{Q}^\dagger\rbrace}  
v_1^{\frac{R_1}{2}}v_2^{\frac{R_2}{2}} v_3^{\frac{R_3}{2}} \,
p^{J_1}q^{J_2}\,
\right]\,.
\end{equation}
It captures the $\frac{1}{16}$-BPS sector, on which
\begin{equation}
\lbrace Q,Q^\dagger\rbrace=E-J_1-J_2-\tfrac{1}{2}(R_1+R_2+R_3) =0\,.
\end{equation}
The charges $J_{1,2}=J_L\pm J_R$ are the Cartan generators of the $SO(2)_L\times SO(2)_R\subset SO(4)$ that parameterizes the angular momenta along $S^3$, and
$R_{1,2,3}$ are Cartan generators of the~$U(1)^3 \subset SU(4)_R$ that parameterize the R-symmetries; the fugacities corresponding to $J_{1,2}$ and $R_{1,2,3}$ are denoted by
\begin{equation}
p=\te(\sigma)\,,\quad q=\te(\tau)\,, \quad 
v_{1,2,3}=\te(\Delta_{1,2,3})\,,
\end{equation}
and obey the constraints
\begin{equation}
\sum^3_{a=1}\Delta_{a}  =\tau+\sigma
\qquad\textrm{or equivalently}\qquad
\prod^3_{a=1}v_{a}=pq 
\,.  
\end{equation}
The $U(N)$ superconformal index can be computed by the integral~\cite{Romelsberger:2005eg,Kinney:2005ej,Dolan:2008qi}
\begin{equation}\label{eq:indexUN}
\mathcal{I}(v_{1,2};p,q)\,=\,{\kappa}\oint_{|w_i|=1} \,\prod_{i=1}^N  \frac{dw_i}{2\pi \text{i} w_i}\,\cdot\, I \,,
\end{equation}
where the integrand is a product of multiple copies of a single variable function $\mathcal{R}_{p,q}$: 
\begin{equation}
I:= \prod_{i<j} \mathcal{R}_{p,q}(w_{ij})\,,
\end{equation}
with variables
\begin{equation}
w_{ij}\,\equiv\,
\frac{w_{i}}{w_{j}}\,.    
\end{equation}
The function $\mathcal{R}_{p,q}$ is defined as
\begin{equation} 
\begin{split}
\mathcal{R}_{p,q}(w) &\,:=\, \frac{\prod^3_{a=1}\Gamma_{\pm}(w {v}_a) }{\Gamma_{\pm}(w)} \,,
\end{split}
\end{equation}
where
\begin{equation}
\begin{split}
\Gamma_{\pm}(w {v}_a) &\,:=\,\Gamma(w {v}_a;p,q)\, \Gamma(\frac{{v}_a}{w};p,q)
 \end{split}
\end{equation}
is a product of two elliptic Gamma functions $\Gamma$~(defined in eq.~\eqref{GammaeDef}).
Finally, the prefactor $\kappa$ is defined as $\kappa:=\kappa_{G= U(N)}$ with
\begin{equation}
\kappa_G \,:=\, \frac{ \Bigl((p; p)_{\infty} (q; q)_{\infty}\prod_{a=1}^3\Gamma_e(v_a;p,q)\Bigr)^{\text{rk}(G)}}{|\mathcal{W}_G|}\,.
\end{equation}

\subsection{Case with $p=q$}
\label{ssec:pq}

We first focus on the case with equal rapidities 
\begin{equation}
p=q=\te(\tau) \,,   
\end{equation}
and define $ \mathcal{R}:= \mathcal{R}_{q,q}$.
Sometimes it is convenient to use the rescaled rapidities 
\begin{equation}
Y_a:=\frac{{v}_a}{\widetilde{q}^2}\,,
\end{equation}
with $\widetilde{q}$ defined as $q=\widetilde{q}^3   \,.$
$Y_a$ satisfy the balancing constraint
\begin{equation}
Y_1 Y_2 Y_3 \,=\,1\,.   
\end{equation}
Without loss of generality we will assume from now on that~$q=|q|\,<\,1\,$.

\medskip

Although there are $N$ gauge rapidities $w_i$, with $i=1,\ldots N$, only their ratios $\frac{w_i}{w_j}$ enter the integrand.
As a result, there are only $N-1$ independent variables, which we choose to be 
\begin{equation}
\underline{z}\,:=\{z_{i}=\frac{w_1}{w_{i+1}}\} 
\qquad \textrm{with} 
\quad i=1,2,\dots, N-1\,.
\end{equation}
In terms of these variables the ratios $\frac{w_i}{w_j}$ are
\begin{equation}\label{eq:AffineChange}
w_{ij}\,:=\,
\frac{w_i}{w_j}\,=\,\begin{cases}
\frac{z_{j-1}}{z_{i-1}} & \qquad i,j \,\geq\, 2\,,\\ {z_{j-1}} & \qquad i\,=\,1\,,\, j\,\geq\, 2 \\ \frac{1}{z_{i-1}} &\qquad i\,\geq \, 2\,, \, j\,=\,1\,.
\end{cases}
\end{equation}
We will call $\{z_i\}$ \emph{affine variables}.
(Note that we will still present some of the expressions in terms of the $w_{1,\dots, N}$ variables to make the Weyl group symmetry $S_N$ more transparent; but one should always keep in mind that there are only $N-1$ independent variables, represented by $z_i$ for example, that enter the Bethe ansatz equation.)

Using the series of identities described in the intermediate equalities below, the $U(N)$ index in eq.~\eqref{eq:indexUN} can be rewritten in terms of affine variables: 
\begin{equation}
\begin{split}
\mathcal{I} &\,=\,\underbrace{\biggl({\kappa}_{U(1)}\,\oint_{|w_1|=1} \prod_{i=1}^N \frac{d w_i}{2\pi \text{i} w_i}\,1\biggr)}_{\mathcal{I}_{U(1)}}
\,\times\, \underbrace{\biggl(\kappa_{SU(N)}\,\oint_{|w_i|=1} \prod_{i=2}^N \frac{d w_i}{2\pi \text{i} w_i}\, I}_{\mathcal{I}_{SU(N)}}\biggr) \\ 
&\,=\,
    \mathcal{I}_{U(1)} \,\times\,\underbrace{{\kappa_{SU(N)}\oint_{|z_i|=1} \prod_{i=1}^{N-1} \frac{d z_i}{2\pi \text{i} z_i}\, \,I}}_{\mathcal{I}_{SU(N)}}
\,=\,\oint_{|z_i|\,=\,1} \prod_{i=1}^{N-1} \frac{d z_i}{2\pi \text{i}z_i}\,\cdot\, {\kappa}I\,.
\end{split}
\end{equation}
The function $\mathcal{R}$ satisfies a quasi-periodicity property
\begin{equation}
\mathcal{R}(\tfrac{z}{q})\,=\, \mathcal{R}(z) \mathcal{S}(z)\,, \qquad \mathcal{R}\left({z}{q}\right)\,=\, \mathcal{R}(z) \mathcal{S}^{-1}(z)
\end{equation}
where the quasi-peridicity $\mathcal{Q}$ is given by the scattering function $\mathcal{S}$, defined as
\begin{equation}
\mathcal{S}(z):=\mathcal{S}(z;q)\,:=\mathcal{S}(z,v_a;q)\,:=\frac{\theta_0(\frac{z}{v_1};q)}{\theta_0({z}{v_1};q)}\frac{\theta_0(\frac{z}{v_2};q)}{\theta_0({z}{v_2};q)}\frac{\theta_0({z} v_1 v_2;q)}{\theta_0(\frac{z}{v_{1}v_2};q)}\,,
\end{equation}
where $\theta_0$ is the Jacobi theta function defined in~\eqref{ThetaDef}, and we have used
\be\label{eq:QuasiPGammas}
\G(p\z;p,q) \= \theta_0(\z;q)\G(\z;p,q)\,, \qquad  \G(q\z;p,q) \= \theta_0(\z;p)\G(\z;p,q)\,.
\ee
The function $\mathcal{S}$ has the following properties
\be\label{eq:symmetries}
\begin{split}
\mathcal{S}(1)&\,=\,-1\,, \qquad
\mathcal{S}(\tfrac{1}{z})\,=\,\mathcal{S}(z)^{-1} \,,\\ \mathcal{S}(z;q)&=\mathcal{S}(z \text{e}(1);q)=\mathcal{S}(z q;q) \\ &=\mathcal{S}(z;q \text{e}(1))\,=\,\mathcal{S}\left(\text{e}\Bigl(\tfrac{\log z}{2\pi\text{i}\tau}\Bigr),\text{e}\Bigl(\tfrac{\Delta_a}{\tau}\Bigr);\text{e}\Bigl(-\tfrac{1}{\tau}\Bigr)\right)\,.
\end{split}
\ee
More generally, it is modular invariant
\begin{equation}
\mathcal{S}(z;q)\,=\,\mathcal{S}\left(\text{e}\Bigl(\tfrac{\log z}{2\pi \text{i}\bigl(c\tau+d\bigr)}\Bigr),\text{e}\Bigl(\tfrac{\Delta_a}{c\tau+d}\Bigr);\text{e}\Bigl(\tfrac{a \tau+b}{c\tau +d}\Bigr)\right)\,,
\end{equation}
where~$a,b,c,d$ are integer numbers obeying the condition~$ad - bc=1$. In particular,~\eqref{eq:symmetries} imply
\be\label{IdentityN2}
\mathcal{S}(\text{e}\bigl(\pm\tfrac{1}{2}\bigr))\,=\,\mathcal{S}(\text{e}\bigl(\pm\tfrac{\tau}{2})\bigr)\,=\,\mathcal{S}(\text{e}\bigl(\pm\tfrac{\tau\pm 1}{2}\bigr))\,=\,1\,,
\ee
or
equivalently
\be
\mathcal{S}(-1)\,=\,\mathcal{S}(\sqrt{q})\,=\,\mathcal{S}(-\sqrt{q})\,=\,1\,.
\ee
This will be useful below.

\subsection{Bethe expansion}\label{sec:3p3}

Given the conditions explained in section~\ref{sec:2}, and one such meromorphic function $I\,$, the choice of
\be
\underline{\mathcal{Q}}\,=\,\{\mathcal{Q}_i\}
\qquad\textrm{with}\quad
i\,=\, 1\,\ldots\, N-1\,,
\ee
which corresponds to a generic choice of vectors~$\underline{n}^{(j)}$'s in definition~\eqref{eq:ShiftOperators},
defines the middle-dimensional contour~$\mathcal{C}^{\underline{\underline{n}}}\,$, and a Bethe formula
\be\label{BA}
\mathcal{I}\,=\, \int_{\mathcal{C}^{\underline{\underline{n}}}} \prod_{i=1}^{N-1} dz_i\, \cdot I_{\textrm{\textrm{B.A}}} \,=\, \int_{\mathcal{C}^{\underline{\underline{n}}}} \prod_{i=1}^{N-1}\frac{d z_i}{z_i}\cdot \frac{\kappa I}{\prod_{i=1}^{N-1} (\mathcal{Q}^{\underline{\underline{n}}}_i-1)}\,.
\ee
For example, we can choose\footnote{
Note that it is a different choice than the one in~\eqref{eq:ShiftOperators}.}
\be\label{eq:QuasiPChoice}
\mathcal{Q}^{\underline{\underline{n}}}_{i}\,=\, Q_{i}
\qquad\textrm{with}\quad
i\,=\, 1\,\ldots\, N-1\,,
\ee
where 
\be\label{eq:BASOps}
{Q}_i\,=\,\prod_{j\neq i=1}^{N}\Bigl(\mathcal{S}(w_{ij})\Bigr)^{s}\,,\qquad s\,=\,\pm 1\,,
\ee
are the $U(N)$ Bethe operators. For completeness, we compare our definition with the $U(N)$ Bethe operator defined in eq.\ (2.16) of~\cite{Benini:2021ano}, which we denote as $Q^{\text{there}}_i\,$. 
With the following identification of variables 
\be
w^{\text{here}}_{ij}\,=e(u^{\text{there}}_{ij})\,,\quad 
\Delta^{\text{here}}_a\,=\,\Delta^{\text{there}}_{a}
\,,\quad
\tau^{\text{here}}\,=\, \tau^{\text{there}}\,,
\ee
we have
\be\label{eq:TranslationToBetheOperator}
Q^{\text{here}}_i\,=(-1)^{N-1}\,\Bigl(Q^{ \text{there}}_i\Bigr)^s\,.
\ee
We recall that in~\eqref{eq:BASOps} and forthcoming definitions, the $w_{ij}$'s are defined as the functions of the $z_i$'s indicated in~\eqref{eq:AffineChange}.
Note that
 \begin{equation}
 Q_N\,=\, \frac{1}{Q_1 Q_2\ldots Q_{N-1}}\,. 
 \end{equation}

\medskip
 
The quasi-periodicity factors~$Q_i$ defined in~\eqref{eq:BASOps} are induced on~$I$ by the shifts
\begin{equation}
w_{i}\to \frac{w_{i}}{q^s}
\qquad\textrm{with}\quad
i\,=\, 1,2,\dots, N-1\,.   
\end{equation}
In particular, the shift $i=1$ in variables $w$ corresponds to $N-1$ simultaneous shifts in affine variables $z$
\be
z_i\,\to\, \frac{z_i}{q^s}
\qquad\textrm{with}\quad
i\,=\, 1,2,\dots, N-1\,.
\ee
The shift $i\neq1$ in a single variable $w$ corresponds to the shifts in the single affine variable $z_{i-1}$:
\be
z_{i-1}\,\to\,q^s \, z_{i-1}\,.
\ee
For example at $N=3\,$, this corresponds to the choice of shift operators $\widehat{\mathcal{Q}}_j$ defined by the vectors
\be
\begin{split}
\underline{n}^{(j=1)}&\,=\,s\{1,1\}\,,  \qquad \underline{n}^{(j=2)}\,=\,s\{0,-1\}\,,
\end{split}
\ee
or equivalently 
\be
\begin{split}
\underline{\underline{n}}&\,=\,s\,\Bigl(\begin{array}{cc} 1 & 1 \\ 0 & -1 \end{array}\Bigr) \,\in\, U^*(2,\mathbb{Z})\,.
\end{split}
\ee
Then, as previously explained, the Bethe contour $\mathcal{C}^{\underline{\underline{n}}}\,$ 
is fixed to be
\be
\mathcal{C}^{\underline{\underline{n}}}\,=\, ({\boldsymbol{\ast}})_{j=1}^{N-1}(\mathcal{C}^{\underline{n}^{(j)}}_j-\mathcal{C}_0)\,=:\,\partial \mathcal{A}^{\underline{\underline{n}}}(q)\,,\qquad \mathcal{C}^{\underline{\underline{n}}}_j = \widehat{\mathcal{Q}}^{\underline{\underline{n}}}_j\,\star\,\mathcal{C}_0\,
\ee
where we recall that
\be
\begin{split}
\mathcal{C}_0&\,:=\,\{{z}_i\,:\,|z_1|=|z_2|=1\}\,.
\end{split}
\ee
From~\eqref{eq:ContourChange} it follows that $\mathcal{C}^{\underline{\underline{n}}}$ can be substituted by  $\mathcal{C}\,$, the boundary of the annulus
\be\label{eq:annulus}
\mathcal{A}(q)\,:=\,\left\{z_{i}: 1 \leq |z_i| \leq \frac{1}{q}\right\}\,.
\ee
(independently of the value of $s$ which from now on we choose to fix $s=-1$, without loss of generality). This is the contour originally used in~\cite{Benini:2018mlo}\cite{Lezcano:2021qbj,Benini:2021ano}. 

\medskip

In concrete checks, where multivariate residue ambiguity can happen, e.g.\,, for $N\geq 3\,$, the geometry of $\mathcal{C}$ defines a set of constraints among the integers~$\mathfrak{m}_{\underline{z}^{*}}$'s. The remaining unconstrained~$\mathfrak{m}_{\underline{z}^{*}}$'s parameterize the ambiguity in selecting multivariate residues, given the fixed contour $\mathcal{C}$. 
The dependence on the unconstrained $\mathfrak{m}_{\underline{z}^{*}}\,$'s disappears after summing over Bethe solutions. 
The constraints that $\mathcal{C}\,$ imposes among the~$\mathfrak{m}_{\underline{z}^{*}}$'s will be algebraically determined by demanding the equality \eqref{BA} to hold, up to the highest possible order in the small-$\widetilde{q}$ expansion.

\subsection{Bethe expansion for $p\,\neq\, q$}
\label{sec:3.4}

Let us consider the $U(N)$ index at different $p$ and $q\,$'s~\cite{Aharony:2024ntg}. 
We also choose 
\be
q\,=\, t^{\frac{\ell_q}{\ell_{t_q}}},\qquad   p \,=\, t\,, 
\ee
which corresponds to the periodization
\be
z_{i}\,\sim\, \frac{z_{i}}{p^{n^{(j)}_i}}
\qquad\textrm{with}\quad
i\,=\, 1,2,\dots, N-1\,, 
\ee
with the ``$SU(N)$ choice"
\be
n^{(j)}_i\,=\ \delta_{i,j}\,.
\ee
More general cases with $\ell_{p}, \ell_{t_{p,q}}\neq 1$ can be studied analogously.

\bigskip

Generalizing the computation for the $p=q$ case, we obtain
\be
\frac{\mathcal{R}_{p,q}[\frac{z}{p}]}{\mathcal{R}_{p,q}[{z}]}\,:=\, \mathcal{S}_{p,q}[z]=\,\frac{\theta_0(\frac{v_1}{z};q)\theta_0(\frac{v_2}{z};q)\theta_0(\frac{v_3}{z};q)\theta_0(\frac{v_1 v_2 v_3}{z};q)}{\theta_0(\frac{z v_1}{p};q)\theta_0(\frac{z v_2}{p};q)\theta_0(\frac{z v_3}{p};q)\theta_0(\frac{z v_1 v_2 v_3}{p};q)} 
\ee
where the new scattering function $\mathcal{S}_{p,q}(z)$ obeys the relations
\be
\mathcal{S}_{p,q}(\frac{p}{z})\,=\,\mathcal{S}_{p,q}(z)^{-1} \,,\quad
\mathcal{S}_{p,q}(z)\,=\,\mathcal{S}_{p,q}(z \text{e}(1))\,=\,\mathcal{S}_{p,q}(z q)\,, \quad \mathcal{S}_{q,q}(z)\,=\, \mathcal{S}(z)\,.
\ee
Using this formula we compute the $SU(N)$ Bethe discontinuities. They are
\be
\frac{I[\underline{z}]\Bigl|_{z_i\to \frac{z_i}{p}}}{I[\underline{z}]}\,=\,\mathcal{Q}_i(\underline{z}) :=\mathcal{S}_{p,q}(z_i)\,\prod_{ j\,\neq\, i\,=\,1}^{N-1}\, \mathcal{S}_{p,q}(\tfrac{z_{i}}{z_j})\,.
\ee
In Appendix~\ref{sec:PneqQ} we will check the case with $N=2$, namely that the $U(2)$ index with $p=q^2\neq q$ is reproduced by the formula
\be
\begin{split}
\mathcal{I}&\,=\,\sum_{{z}_1^\ast\,\in\,\mathcal{A}} \text{Res}\Biggl[\frac{{I}[{z}_1]}{ z_1( \mathcal{Q}_1-1)}\,;\, {z}_1^\ast\Biggr] 
\,=\,\sum_{{z}_1^\ast\,\in\,\mathcal{A}(p)} \frac{I[z_1^\ast]}{\Biggl( z_1 \frac{\partial\mathcal{Q}_1}{\partial z_1}\Biggr)\Biggr|_{{z}_1=z_1^\ast}}\,,
\end{split}
\ee
with 
\be
\mathcal{S}_{p,q}(z_1^\ast)\,=\,1\,,
\ee
and $z_1^\ast$ lying in the two-dimensional annulus $\mathcal{A}(p)\,$.

\section{Classification of Bethe solutions}\label{sec:4}

Let us come back to the particular case with $p=q\,$. 
In this case, the chosen $U(N)$ Bethe equations are
\begin{equation}\label{BEqs}
{Q}^{-1}_i(\underline{z})\,=\,\prod_{j\,\neq\, i\,=1}^{N}\mathcal{S}(w_{ij})\,=\,1
\qquad\textrm{with}\quad
i\,=\, 1,2,\dots, N-1\,, 
\end{equation}
where the $w_{ij}=w_{ij}(\underline{z})$ were defined in~\eqref{eq:AffineChange}.

Let us consider the symmetry of the set of Bethe equations~\eqref{BEqs}.
First of all, the individual quasi-periodicities $Q_i$ are invariant under the lattice-shifts
\be\label{eq:aLatticeIdent}
z_i \,\sim\, z_i q =: z_i \widetilde{q}^3\,.
\ee
Consider the cyclic subgroup $\mathbb{Z}_N$ of the permutation group $S_N$, which is the Weyl group of $U(N)$.

This set of equations is invariant under the $N-1$ transformations
($k=1\,, \dots\,,\, N-1 $) 
\be
\Xi_{k}:\qquad z_{i} \mapsto \begin{cases} \frac{1}{z_{i}} & \text{if} \qquad i \,=\,k\\ \frac{z_{i}}{z_{k}}& \text{if} \qquad i\,\neq\, k\end{cases} \,.
\ee

The Bethe roots are \emph{degenerate} (resp.\ \emph{non-degenerate}), if their Jacobian $H=0$ (resp.\ $H\neq 0$).

\subsection{Topological solutions from symmetry}  \label{sec:TopSols}

Topological solutions to~\eqref{BEqs} are defined as solutions which are left invariant by the operations~$\Xi_k\,$ under the lattice identifications~\eqref{eq:aLatticeIdent}. 
They can be classified as \emph{irreducible} or \emph{reducible}.

\begin{definition}
\emph{Irreducible topological solutions} are topological solutions which are by construction non-degenerate.
They can always be written in the form
\be\label{eq:IrreducibleTopSols}
\boldsymbol{z}^*_i\,=\,
\text{e}(u_{1,1+i})\,,\qquad u_{1,1+i} \,=\, \frac{T}{N} i\,
\ee
with $T$ being periods of the lattice~$\mathbb{Z}+\tau \mathbb{Z}\,$ such that all $u_{1,1+i}$ are different.\footnote{
These are the Hong-Liu solutions~\cite{Benini:2018ywd,Hong:2018viz}. The are also the $(m,n)$-fixed points of~\cite{Cabo-Bizet:2019eaf,Cabo-Bizet:2020nkr,Cabo-Bizet:2020ewf}.  Their positions are determined in terms of the lattice symmetry $z_i \sim z_i q$ and the cyclic operations $\Xi_k$ defined in~\eqref{eq:symmetries}~\cite{Cabo-Bizet:2019eaf}.}
\end{definition}

\begin{definition}
\emph{Reducible topological solutions} are solutions for gauge group~$U(N)$ obtained by assembling topological solutions of Bethe equations for gauge group $U(N^\prime)$ with $N^\prime< N\,$.\footnote{
The assembled solutions are also preserved by the operations~$\Xi_k\,$ at rank $N^\prime\,$.} These solutions exist only at $N\geq 3$. In the cases we have studied they are intersected by continuous solutions. Thus, they happen to be degenerate. 
\end{definition}

\subsubsection{Topological solutions for $N=2,3,4,5$}

Let us now give some examples of topological solutions with small $N$.
\begin{itemize}
\item $N=2$. For $U(2)$, there is a single gauge rapidity $z_1$ and a single $U(2)$ Bethe equation
\be
Q_1(z_1)\,=\,1\,,
\ee
which can be rewritten as
\be
\begin{split}
\mathcal{S}(z_1)&\,=\,1 \,.
\end{split}
\ee
Then from identities~\eqref{IdentityN2} there follow, trivially, the $3$ irreducible topological solutions
\be\label{eq:U2sols}
\{\boldsymbol{z}_1^{\ast}\}=
\Bigl\{\pm 1\,,\,\pm\sqrt{q}\,\Bigr\}\,.
\ee
The $+1$ choice is actually trivial because the integrand of the index vanishes at it.

\item $N=3$. In this case there are two rapidities $\{z_1,z_2\}$ and two $U(3)$ Bethe equations:
\be
Q_1(\underline{z})=1 \,,\qquad Q_2(\underline{z}) =1\,,
\ee
or equivalently, in terms of the scattering function,
\be
\begin{split}
\mathcal{S}(z_1)\mathcal{S}(z_2)=1 \,,\,\qquad \mathcal{S}(\tfrac{1}{z_1})\mathcal{S}(\tfrac{z_2}{z_1})&\,=\,1\,.
\end{split}
\ee
Note that the second equation is obtained from the first after applying the operation $\Xi_{1}\,$, and \emph{viceversa}.
The irreducible topological solutions are well known. They take the from~\eqref{eq:IrreducibleTopSols}. Reducible topological solutions of rank ${N}^\prime-1=1$ are:
\be\label{eq:ReducibleN3}
\begin{split}
\{\{\mathfrak{z}^*_1,\mathfrak{z}^*_2\}\} =&\Bigl\{\{-1,\sqrt{q}\},\{-1,-\sqrt{q}\},\{\sqrt{q},-\sqrt{q}\}\Bigr\}\,\\  \cup\, &\Bigl\{\{\sqrt{q},-1\},\{-\sqrt{q},-1\},\{-\sqrt{q},\sqrt{q}\}\Bigr\} 
\end{split}
\ee
All these 6 solutions will happen to be intersected by continuous solutions.

\item $N=4$. In this case there are three rapidities $\{z_{1},z_{2},z_{3}\}$ and three $U(4)$ Bethe equations:
\be
Q_1(\underline{z})=1 \,,\qquad Q_2(\underline{z}) =1\,,\qquad
Q_3(\underline{z}) =1\,,
\ee
or equivalently, in terms of the scattering function,
\be
\begin{split}
\mathcal{S}(z_1)\mathcal{S}(z_2)\mathcal{S}(z_3)\,=\,1 \,,\, \quad 
\mathcal{S}(\tfrac{1}{z_1})\mathcal{S}(\tfrac{z_2}{z_1})\mathcal{S}(\tfrac{z_3}{z_1})\,=\,1\,,\quad
\mathcal{S}(\tfrac{z_1}{z_2})\mathcal{S}(\tfrac{1}{z_2})\mathcal{S}(\tfrac{z_3}{z_2})\,=\,1\,,
\end{split}
\ee

Irreducible topological solutions are given by~\eqref{eq:IrreducibleTopSols}. An example of reducible topological solution of rank $1$ is
\be
\{\mathfrak{z}^*_1,\mathfrak{z}^*_2,\mathfrak{z}^*_3\} =\Bigl\{-1,\sqrt{q}, -\sqrt{q}\Bigr\}\,.
\ee

\item $N=5$.  In this case there are four rapidities $\{z_1,z_2,z_3,z_4\}$
and four $U(5)$ Bethe equations:
\be
Q_1(\underline{z})=1 \,,\, Q_2(\underline{z}) =1\,,\, Q_3(\underline{z}) =1\,,\, Q_4(\underline{z}) =1\,,
\ee
or equivalently, in terms of the scattering function,
\be
\begin{split}
\mathcal{S}(z_1)\mathcal{S}(z_2)\mathcal{S}(z_3)\mathcal{S}(z_4)&\,=\,1 \,,\, \\ \mathcal{S}(\tfrac{1}{z_1})\mathcal{S}(\tfrac{z_2}{z_1})\mathcal{S}(\tfrac{z_3}{z_1})\mathcal{S}(\tfrac{z_4}{z_1})&\,=\,1\,,\\
\mathcal{S}(\tfrac{z_1}{z_2})\mathcal{S}(\tfrac{1}{z_2})\mathcal{S}(\tfrac{z_3}{z_2})\mathcal{S}(\tfrac{z_4}{z_2})&\,=\,1\,, \\ 
\mathcal{S}(\tfrac{z_2}{z_3})\mathcal{S}(\tfrac{z_1}{z_3})\mathcal{S}(\tfrac{1}{z_3})\mathcal{S}(\tfrac{z_4}{z_3})&\,=\,1\,,
\end{split}
\ee

An example of reducible topological solutions of rank $(3,3)$. This notation means that one can assemble two irreducible solutions at rank $N^\prime=3$ to construct a solution for the $U(5)$ system. This solution is
\be
\{\mathfrak{z}^*_1,\mathfrak{z}^*_2,\mathfrak{z}^*_3,\mathfrak{z}^*_4\} =\Bigl\{\text{e}(\tfrac{1}{3}),\text{e}(\tfrac{2}{3}), \text{e}(\tfrac{\tau}{3}),\text{e}(\tfrac{2\tau}{3})\Bigr\}
\ee
There are other obvious permutations and generalizations. However, a more detailed study of those will be presented elsewhere. As it has been just illustrated, using the symmetry $\Xi_k$ one can classify all topological solutions at generic~$N\,$.
\end{itemize}

\subsubsection{Topological solutions for the index with $p\neq q$.}\label{subsec:TopPNeqQ}
It is actually very easy to generalize the previous analysis, which assumed $p=q$, to the case $p \neq q\,$. In the latter, for a $U(N)$ theory, we have chosen to focus on the $SU(N)$ Bethe equations
\be
\mathcal{Q}_i(\underline{z})\,=\,1\,, \qquad i\,=\,1\,,\,\ldots\,, N-1\,,
\ee
defined by ($t=p$) with
\be\label{eq:QPQOperator}
\mathcal{Q}_i(\underline{z})= {Q}^{(p,q)}_i :=\mathcal{S}_{p,q}(z_i)\,\prod_{ j\,\neq\, i\,=\,1}^{N-1}\, \mathcal{S}_{p,q}(\tfrac{z_{i}}{z_j})\,,\qquad i\,=\, 1\,,\,\ldots\,, \,N-1\,.
\ee

\begin{theorem}\label{th:4p1}
If $\underline{\zeta}$ is a topological solution of the $p=q$ $U(N)$ Bethe equations
\be
Q_i(\underline{\zeta})\,=\,1 \,,
\ee
then for any $p$ and $q$
\be
\underline{z}\,=\, \sqrt{p} \zeta
\ee
is a solution of the generalized $SU(N)$ Bethe equations:
\be
Q^{(p,q)}_i(\sqrt{p}\zeta)\,=\,1\,.
\ee
\end{theorem}
For irreducible topological roots, this conclusion follows from the symmetry properties
\be
S_{p,q}(\sqrt{p} \tfrac{1}{z})\,=\,\frac{1}{S_{p,q}(\sqrt{p} z)}
\ee
the double periodicity
\be
S_{p,q}(\sqrt{p} {z}) \,=\,S_{p,q}(\sqrt{p} q {z})\,=\, S_{p,q}(\sqrt{p} \text{e}(1) {z})
\ee
and the symmetries $\Xi_k\,$, or equivalently, from the Lemma noted in~\cite{Cabo-Bizet:2019eaf}.\footnote{
For non-irreducible ones, the observation above follows trivially from the latter reasons together with the fact that the spectrum of gauge charges $\rho$ is the same for the cases $p=q$ and $p\neq q\,$.}

For example, for generic $p\neq q$ there are four seed solutions to the $SU(2)$ Bethe equation
\be
\mathcal{S}_{p,q}(x)\,=\,1\,.
\ee
These are
\be\label{eq:U2solsGenCase}
\{x\}\,=\,\{\,\pm \sqrt{p}\,,\, \pm \sqrt{p q}\}\,.
\ee
They follow from observation~\ref{th:4p1} and equation~\eqref{eq:U2sols}. New solutions can then be generated by successive action of $q$-shifts $x\to q x\,$. One must then consider contributions from all $q$-shifted images lying within the annulus $\mathcal{A}(p)$\,.

\subsection{Non-topological solutions}\label{subsec:NonTopSolsDef}

Not all solutions that contribute to the index are topological. 
However, irrespective of whether the Bethe roots are topological or not, all the solutions we have found obey the following asymptotic \emph{boundary conditions}:
\be\label{eq:AsymtoticCond}
z_i\,\underset{\widetilde{q}\to 0}{=}\, \mathcal{O}(\widetilde{q}^0)\,\times\,(\text{topological solutions})\,.
\ee
We conjecture that the boundary condition \eqref{eq:AsymtoticCond} holds for all solutions, based on the following two arguments:
\begin{enumerate}
\item We have reduced the Bethe equations around $\widetilde{q}=0$ for a vast number of generic ansatze not of the form \eqref{eq:AsymtoticCond}; but all these more generic ansatze lead to equations with no solution. 
\item As we will show later, the solutions that we find using this ansatz collectively reproduce the correct index, to the order we check.\footnote{Although there is a slight possibility that there exist additional Bethe solutions not captured by the ansatz (4.26), because all these additional solutions must then collectively contribute zero to the index (which is an infinite series), we think this is a highly unlikely scenario, in particular given the fact that more generic ansatze haven't lead to any solutions.}  
\end{enumerate}
It would be interesting to try to check or prove this conjecture.


\smallskip

Once all the topological solutions are identified,~\eqref{eq:AsymtoticCond} leads to a systematic identification of all contributing Bethe roots, which we move on to explain below.

The condition ~\eqref{eq:AsymtoticCond} can be stated concretely as: 
\be\label{eq:BCBetheSharpen}
\begin{split}
z_i& = Z_i\, {\Bigl(\widetilde{q}^3\Bigr)}^{\frac{ c_i}{N^\prime} i}\,,\qquad i\,=\,1\,,\ldots\,, N-1\,,
\end{split}
\ee
where the integers
\be
\begin{split}
c_{i}&\,\sim\, c_{i}+N^\prime\,,
\end{split}
\ee
define a topological solution of rank $N^\prime=2, 3\,,\ldots,\,N \,$. In the terminology introduced in section~\ref{eq:RecursiveMethod} the primary spectrum of Bethe solutions is
\be
\underline{\Delta}\,=\,\Bigl\{\Delta_i = 1-\frac{c_i }{N^\prime}i+\lfloor \frac{c_i }{N^\prime} i\rfloor\Bigr\}\,.
\ee
The boundary conditions~\eqref{eq:BCBetheSharpen} that correspond to irreducible topological solutions will be called \emph{irreducible boundary conditions}. Those that correspond to reducible topological solutions will be called \emph{reducible boundary conditions}. The total set of boundary conditions in the case $N=3$ has been summarized in table~\ref{table:N_prime_values_filtered}.

The unknown function
\be
Z_i= Z_i[\widetilde{q}]
\ee
is regular at $\widetilde{q}=0\,$. It has the Taylor expansion
\be
Z_i\,=\,\sum_{n=0}^{\infty} Z^{(n)}_{i} \widetilde{q}^{\,\kappa n }\,,\qquad  \kappa  := \frac{3}{N^\prime}-\lfloor\frac{3}{N^\prime}\rfloor\,. 
\ee
Plugging the ansatze~\eqref{eq:BCBetheSharpen} in the Bethe equations
\be\label{BetheEqs}
Q_i(\underline{z})=1\,, \qquad i\,=\,1\,, \,\ldots\,  N-1\,,
\ee
and perturbatively expanding the result in the left-hand side around $\widetilde{q}\,=\,0\,$, one can then solve an iterative linear system for infinitely many Taylor coefficients of the $Z_i$'s in terms of $Z^{(0)}_i$, $Y_1$ and $Y_2\,$.  

Using this method, we have recovered all possible solutions for $N=3\,$. 
We think that this method can give all the solutions for any $N\,$. The reason being the following. The contribution of the topological roots to the superconformal index produces tachyonic singularities in the expansion $\widetilde{q}\to0$~\cite{Lezcano:2021qbj,Benini:2021ano}. 
In order to cancel such tachyonic contributions, the continuous solutions must obey the very same boundary conditions dictated by topological solutions in the expansion $\widetilde{q}\to0\,$. Indeed, for $N=3$, the continuous solutions we have found \emph{via} this method cancel all tachyonic contributions in a rather non-trivial way, as it should be. This will be shown in the following section.

\subsection{A solution-generating procedure} \label{subsec:4.3}

Let us illustrate some cases of boundary conditions, or equivalently primary weights $\underline{\Delta}\,$, and their corresponding solutions. A systematic analysis will be postponed until section~\ref{sec:5} where the proposed methods will be tested. Examples are numerated as in the second column of table~\ref{table:N_prime_values_filtered}

\begin{definition}The \emph{irreducible boundary condition 1 (BC1)} is
\be
N^\prime=N=3\,, \, c_i=0\,,
\ee
which corresponds to 2 out of the possible 8 irreducible topological roots and defines the ansatz
\be
\begin{split}
z_1&\,=\, Z_1 \,, \qquad 
z_2\,=\, Z_2 \,,
\end{split}
\ee
with
\be
Z_{i}\,=\, Z_{i}^{(0)}\,+\, \widetilde{q} Z_{i}^{(1)}\,+\,\ldots\,.
\ee
\end{definition}
It is easy to check that there are 3 irreducible topological solutions for which
\be\label{eq:N3DiscreteSols}
\Bigl\{\{Z_1^{(0)},Z_2^{(0)}\}\Bigr\}\,=\,\Bigl\{\{1,1\},\{-(-1)^{\frac{1}{3}},(-1)^{\frac{2}{3}}\},\{(-1)^{\frac{2}{3}},-(-1)^{\frac{1}{3}}\}\Bigr\}
\ee
and all $Z_i^{(j)}$ with $j\,>\,0$ vanish trivially. The first solution we drop, because it corresponds to coincident roots for which the gauge holonomies vanish, $z_1=z_2=1\,$ and thus the supercoformal index vanishes at it. There are no non-topological solutions for these boundary conditions.

\begin{definition} The \emph{irreducible boundary condition 2 (BC2)} is
\be
N^\prime=N=3\,, \, c_i=1\,,
\ee
which corresponds to another 3 out of the 8 irreducible topological roots, and defines the ansatz
\be
\begin{split}
z_1&\,=\, Z_1 \widetilde{q}\,,\, \qquad
z_2\,=\,Z_2 \widetilde{q}^2\,,
\end{split}
\ee
with
\be
Z_{i}\,=\, Z_{i}^{(0)}\,+\, \widetilde{q} Z_{i}^{(1)}\,+\,\ldots\,.
\ee
\end{definition}
It is easy to check that there are 3 irreducible topological solutions for which
\be\label{eq:N3DiscreteSols}
\Bigl\{\{Z_1^{(0)},Z_2^{(0)}\}\Bigr\}\,=\,\Bigl\{\{1,1\},\{-(-1)^{\frac{1}{3}},(-1)^{\frac{2}{3}}\},\{(-1)^{\frac{2}{3}},-(-1)^{\frac{1}{3}}\}\Bigr\}
\ee
and all $Z_i^{(j)}$ with $j\,>\,0$ vanish trivially. For these boundary conditions there are also 2 irreducible non-topological solutions, which at order $\mathcal{O}(\widetilde{q}^0)$ are
\be
\begin{split}
Z^{(0)}_{2}\,=\,Z^{(0)}_{2\mp}\,&\,:=\, \frac{Y_2 Y_1^2 Z^{(0)}_1+Y_2^2 Y_1 Z^{(0)}_1-Y_2 Y_1+Z^{(0)}_1}{2 Y_1
   Y_2 Z^{(0)}_1} \\&\qquad\mp\frac{\sqrt{\left(Y_2 Y_1^2 Z^{(0)}_1+Y_2 Y_1 \left(Y_2 Z^{(0)}_1-1\right)+Z^{(0)}_1\right){}^2-4 Y_1^2 Y_2^2 {Z^{(0)}_1}^3}}{2 Y_1
   Y_2 Z^{(0)}_1}\,.
   \end{split}
\ee
These continuous solutions do not intersect the irreducible topological solutions~\eqref{eq:N3DiscreteSols} at generic $Y_1$ and $Y_2\,$. It does not intersect reducible topological solutions neither.\footnote{
That said, for $Y_2=Y_1\approx 1$ ... these two solutions intersect the $\{1,1\}$ solution in~\eqref{eq:N3DiscreteSols}.}

\begin{definition} The \emph{irreducible boundary condition 3 (BC3)} is
\be
N^\prime=N=3\,, \, c_i=2\,,
\ee
which corresponds to the last 3 out of the 8 possible topological roots. These solutions come from the solutions $3$, $4$, and $5$ above by exchanging the colour indices $1 \leftrightarrow2\,$ in them. Again, these continous solutions do not intersect topological solutions at generic values of $Y_1$ and $Y_2\,$.
\end{definition}

\begin{definition} The \emph{reducible boundary condition 1 (RBC1)} is
\be
N^\prime=N-1=2\,, \, c_1=0\,,\, c_2\,=\,1
\ee
which corresponds to the first 2 reducible discrete roots among the 3 in~\eqref{eq:ReducibleN3} and defines the ansatz
\be
\begin{split}
z_1&\,=\, Z_1 \times (-1) \,, \qquad
z_2\,=\, Z_2 \times \widetilde{q}^{\frac{3}{2}}\,,
\end{split}
\ee
with
\be
Z_{i}\,=\, Z_{i}^{(0)}\,+\, \widetilde{q}^{\frac{1}{2}} Z_{i}^{(1)}\,+\,\ldots\,.
\ee
\end{definition}
It is easy to check that there are 2 reducible topological solutions for which
\be\label{eq:N3DiscreteSolsReducible}
\Bigl\{\{Z_1^{(0)},Z_2^{(0)}\}\Bigr\}\,=\,\Bigl\{\{1,1\},\{1,-1\}\Bigr\}
\ee
and all $Z_i^{(j)}$ with $j\,>\,0$ vanish trivially. These two topological solutions are intersected though by 2 continuous solutions, which at order $\mathcal{O}(\widetilde{q}^0)$ are
\be\label{eq:ReducContSol}
\begin{split}
Z^{(0)}_{2}\,=\,Z^{(0)}_{2\mp}\,&\,:=\,\mp\,\sqrt{Z^{{(0)}}_{1}}\,,
   \end{split}
\ee
and then, they are not independent solutions. Thus, we can safely focus on the two embedding continuous families.

\begin{definition}
\emph{Reducible boundary condition 2 (RBC2)} is
\be
N^\prime=N-1=2\,, \, c_1=1\,,\, c_2\,=\,0
\ee
which corresponds to the permutations of the first 2 reducible discrete roots in~\eqref{eq:ReducibleN3}. Again, these reducible boundary conditions give two continuous solutions which come from solutions~\eqref{eq:ReducContSol} above by exchanging the colour indices $1 \leftrightarrow2\,$. These continous solutions intersect reducible topological solutions.
\end{definition}

\begin{definition}
\emph{Reducible boundary condition 3 (RBC3)} is
\be
N^\prime=N-1=2\,, \, c_1\,=\,c_2\,=\,1\,
\ee
and defines the ansatz
\be
\begin{split}
z_1&\,=\, Z_1 \times \widetilde{q}^{\frac{3}{2}} \,, \qquad 
z_2\,=\,Z_2 \times \widetilde{q}^{\frac{3}{2}}\,,
\end{split}
\ee
with
\be
Z_{i}\,=\, Z_{i}^{(0)}\,+\, \widetilde{q}^{\frac{1}{2}} Z_{i}^{(1)}\,+\,\ldots\,.
\ee
\end{definition}
It is easy to check that there are 2 reducible topological solutions for which
\be\label{eq:N3DiscreteSolsReducible}
\Bigl\{\{Z_1^{(0)},Z_2^{(0)}\}\Bigr\}\,=\,\Bigl\{\{-1,1\},\{1,-1\}\Bigr\}
\ee
and all $Z_i^{(j)}$ with $j\,>\,0$ vanish trivially. These two topological solutions are connected by 1 continuous solution, which at order $\mathcal{O}(\widetilde{q}^0)$ is
\be
\begin{split}
Z^{(0)}_{2}\,&\,:=\,-\,\frac{1}{Z^{{(0)}}_{1}}\,,
   \end{split}
\ee
and then, they are not independent solutions. Thus, we can safely focus on the embedding continuous solution.

\begin{table}[h!]
\centering
\renewcommand{\arraystretch}{1.3} 
\setlength{\tabcolsep}{12pt} 
\begin{tabular}{|c|c|c|}
\hline
\multicolumn{3}{|c|}{\textbf{Topological solutions and boundary conditions for $N=3$}} \\ \hline
\multirow{8}{*}{\textbf{$N^\prime = 3$} \quad (Irreducible)} 
& 
\begin{tabular}{|c|c|}
\hline
\multicolumn{2}{|c|}{\textbf{BCs}} \\   \hline
\textbf{$c_1$} & \textbf{$c_2$} \\ \hline
0 & 0  \\ \hline
1 & 1  \\ \hline
2 & 2 \\ \hline
\multicolumn{2}{|c|}{ $z_i \sim z_i \widetilde{q}^3 = \mathcal{O}(1) \widetilde{q}^{\,c_i i}$}\\\hline
\end{tabular} &
\begin{tabular}{|c|c|c|c|}
\hline
\multicolumn{4}{|c|}{\textbf{Topological solutions}} \\   \hline
\textbf{$c_1$} & \textbf{$c_2$} & \textbf{$d_1$} & \textbf{$d_2$} \\ \hline
0 & 0 & 1 & 1 \\ \hline
0 & 0 & 2 & 2 \\ \hline
1 & 1 & 0 & 0 \\ \hline
2 & 2 & 0 & 0 \\ \hline
1 & 1 & 1 & 1 \\ \hline
1 & 1 & 2 & 2 \\ \hline
2 & 2 & 1 & 1 \\ \hline
2 & 2 & 2 & 2 \\ \hline
\end{tabular}  
\\ 
\hline
\multirow{6}{*}{\textbf{$N^\prime = 2$} \quad (Reducible)} 
& 
\begin{tabular}{|c|c|}
\hline
\textbf{$c_1$} & \textbf{$c_2$} \\ \hline
0 & 1  \\ \hline
1 & 0  \\ \hline
1 & 1 \\ \hline
\multicolumn{2}{|c|}{ $z_i \sim z_i \widetilde{q}^3 = \mathcal{O}(1) \widetilde{q}^{\,\frac{3 c_i }{2} i}$}\\
\hline
\end{tabular}
&
\begin{tabular}{|c|c|c|c|}
\hline
\textbf{$c_1$} & \textbf{$c_2$} & \textbf{$d_1$} & \textbf{$d_2$} \\ \hline
0 & 1 & 1 & 0 \\ \hline
0 & 1 & 1 & 1 \\ \hline
1 & 0 & 0 & 1 \\ \hline
1 & 0 & 1 & 1 \\ \hline
1 & 1 & 1 & 0 \\ \hline
1 & 1 & 0 & 1 \\ \hline
\end{tabular} 
\\ \hline 
\end{tabular}
\caption{The 3 irreducible boundary
conditions (with $N^ \prime= N=3\,$) and the corresponding 8 solutions; and the 3  reducible boundary conditions (with $N^\prime=2\,$) and the corresponding 6 solutions.}
\label{table:N_prime_values_filtered}
\end{table}

\subsection{A convenient alternative procedure}\label{ref:AlgorithmSols}

Instead of solving the Bethe equations it is more efficient to Laurent expand the {Bethe integration form} 
\be
{I}_{\textrm{B.A}} \,{d z_1}\,\wedge\,\ldots\,\wedge\, {dz_{N-1}}\,=\, \frac{\kappa I}{\prod_{i=1}^{N-1}z_i(Q_i-1)}\,{d z_1}\,\wedge\,\ldots\,\wedge\, {dz_{N-1}}
\ee
around every possible topological boundary conditions. We have checked at various $N\geq 2$ that for irreducible topological boundary conditions such expansions have the form\footnote{
For reducible boundary conditions the expansion is different and simpler.}
\be\label{eq:ExpansionBAGeneral}
\frac{\mathcal{K}[\underline{z},q,Y_a] \,{d z_1}\,\wedge\,\ldots\,\wedge\, {dz_{N-1}}}{\Bigl(\prod_{i=1}^{N-2}(Z_i^{(0)N-1}+\prod_{j\neq i\,=\,1}^{N-1}Z^{(0)}_j)\Bigr)\,(1-\prod_{i=1}^{N-1}Z^{(0)}_i)\,\times\, U^{n_N}_{N} \times U^{n_{N+1}}_{N+1}\ldots}\,.
\ee
At leading order in the small-$\widetilde{q}$ expansion, $\mathcal{K}$ is holomorphic in the region surrounded by the integration contour $\mathcal{C}\,$. For $N=3$ $\mathcal{K}$ remains holomorphic at any order in the small-$\widetilde{q}$ expansion. Curiously, at $N\geq4$ the subleading terms come with denominators that involve the very same factors in the denominator in~\eqref{eq:ExpansionBAGeneral} but with increasing integer powers.

The $U_{N}$, $U_{N+1}$, $\ldots$, are a finite number of multivariate polynomials of the $\underline{Z}^{(0)}$'s that depend on the $Y_{a}$'s but not on $\widetilde{q}\,$. The~$n_{N}\,$, $n_{N+1}\,$, $\ldots$, are positive integer numbers. $n_j$ denotes the number of Bethe denominators $Q_i-1$ that share a common factor $U_j$ at leading order in the corresponding  small-$\widetilde{q}$ expansion. It then follows that $n_j\,\leq\, N-1\,$. The zero locus condition of these common factor polynomials $U_{j}$ define continuous solutions of the relevant Bethe ansatz equations. At leading order in the small-$\widetilde{q}$ expansion they correspond to the generalization to $N>3$ of the common factors $h$ discussed around equation~\eqref{eq:Defh}. As we have announced before, only certain discrete points of the continuous solutions contribute to the integration formula.

\medskip

The discrete points $\underline{Z}^{(0)}$ that contribute to the integration formula are simultaneous zeroes of $N-1$ (out of all the possible) denominator-factors that arise from  the expansions~\eqref{eq:ExpansionBAGeneral}. Only those solutions without coincident values contribute. 
The irreducible topological solutions (for the corresponding set of boundary conditions) come from the vanishing of the first $N-1$ factors. 
Judiciously solving $N-2$ out of those vanishing conditions, one is left with a single condition for the remaining variable, which we can take to be $Z_{N-1}^{(0)}\,$, and this remaining condition is 
\be\label{eq:OrderNPolTop}
1\,-\,\Bigl(Z_{N-1}^{(0)}\Bigr)^N\,=\,0\,,
\ee
which is a single variable polynomial of order $N$ and defines $Z_{N-1}^{(0)}$
as $N$-th roots of unity.  Each one of these roots defines one irreducible topological condition consistent with the specified boundary condition $c_i\,$.

For the trivial boundary condition $c_i=0$ ($\underline{\Delta}=0$) the multivariate polynomials $U_{N}\,,$ $U_{N+1}\,$, $\ldots\,$, are trivial (equal to $1$). 
This means that for the corresponding trigonometric-like expansion one only makes contact with $N$ irreducible topological roots, at any $N\,$.

Non-topological solutions depending on $Y_a$ come from the simultaneous vanishing of $n$ out of the $N-1$ polynomials $Y_a$-independent polynomials in the denominator~\eqref{eq:ExpansionBAGeneral} and $N-n-1$ out of the polynomials $U_{N}$, $U_{N+1}$, $\ldots\,$. 

In the following section we will exploit this approach for the case $N=3\,$. Further conclusions on larger values of $N$ will be given elsewhere.

\section{Bethe expansion for the $U(3)$ index at $p=q$}
\label{sec:5}

In this section we explicitly find all the Bethe roots that contribute to the $U(3)$ or $SU(3)$ index $\mathcal{I}$ and evaluate their contributions. 
We demonstrate that the method proposed in previous sections passes a very non-trivial test: the cancellation of tachyonic contributions to $\mathcal{I}$.

We recall that irreducible topological contributions $\underline{z}=\underline{z}^{*}$ are computed by the formula
\be\label{eq:IrreducibleTopRoots}
\mathcal{I}[\underline{z}^{*}] \,=\,\frac{\kappa \,I}{H}\,\Biggl|_{\underline{z}=\underline{z}^{*}}
\ee
where the \emph{one-loop determinant}-like contribution $H$ is the Jacobian~\eqref{eq:Jacobian}. Continuous solutions are degenerate and have vanishing Jacobian $H=0\,$ by definition. Thus, this formula does not work for them. For them one should use instead the approach explained in subsection~\ref{subsec:MethodContinuous} as we proceed to check below.

\subsection{The integrand around topological boundary conditions}

The working example will be $\mathcal{N}=4$ SYM with $U(N=3)$, and flavour rapidities~$Y_2=Y_1\,$. 
In this case the index is
\be\label{IndexToCompare}
\begin{split}
\mathcal{I} &\,=\,1+\widetilde{{q}}^4 \left(3 Y_1^2+\frac{2}{Y_1}+\frac{1}{Y_1^4}\right)
-2 \widetilde{{q}}^5 \left(2
   Y_1+\frac{1}{Y_1^2}\right)+\widetilde{{q}}^6 \left(2
   Y_1^3+\frac{1}{Y_1^6}\right)+\mathcal{O}\left(\widetilde{{q}}^7\right)\,.
   \end{split}
\ee
The goal will be to reach $\mathcal{I}$, up to the highest possible order in powers of $\widetilde{q}\,$, starting from a small $\widetilde{q}$ expansion of
\be
\int_{\mathcal{C}} \prod_{i=1}^{N-1} dz_i\,\cdot\, I_{\textrm{B.A}} \,:=\, \int_{\mathcal{C}} \prod_{i=1}^{N-1}\frac{d z_i}{z_i} \cdot \frac{\kappa I}{\prod_{i=1}^{N-1} (\mathcal{Q}_i-1)}\,,
\ee
around the boundary conditions reported Table~\ref{table:N_prime_values_filtered}, i.e., assuming all possible ansatze (asymptotic behaviours)
\be
\underline{z}= \underline{Z}\, q^{\frac{c_i i}{N^\prime}} \,.
\ee
These boundary conditions take us to the infinitesimal vicinities of all possible Bethe roots. We will assume a generic integration contour $\mathcal{C}^\prime$ within the annulus~\eqref{eq:annulus} and compute all possible Bethe residues within the annulus~\eqref{eq:annulus}
\be
{I}_{\textrm{B.A}} \,=\, \frac{\kappa I}{z_1(1-Q_1)z_2(1-Q_2)}\,{d z_1}\,\wedge\, {dz_2}\,.
\ee
The multivariate residue of such roots will depend on the selection of $\mathcal{C}^\prime\,$. The ambiguity in choice of $\mathcal{C}^\prime$ will be encoded in the choice of $20$ integer numbers.
\be
\begin{split}
n_{6,\dots,10}\,,\,
\qquad 
n_{14,\dots,20}\,,\,  
\qquad
m_{1,\dots,8}\,.
\end{split}
\ee
They label all possible selections of the residues of non-topological roots 
\be\label{eq:NonTopContribRes}
\begin{split}
R_{6,\dots,10}\,,  \, \qquad
R_{14,\dots,28}\,,
\end{split}
\ee
associated to $20$ Bethe roots, which we have labelled by the integer numbers in subscripts. These Bethe roots  are all intersected by continuous Bethe solutions. 
This result is telling us that the contribution of continuous solutions gets localized to certain points on it. 
We will return to this point in more detail later. 

To obtain the full answer, the $20$ non-topological residues must be added to the $8$ irreducible topological residues
\be\label{ResiduesIrredTop}
R_{1,\dots,5}\,,\, R_{11,12,13}\,,
\ee
whose contribution to the Bethe ansatz-type formula we have defined to be~\eqref{eq:IrreducibleTopRoots}.

Once all potentially contributing roots are identified,
the final step is to solve for the values of integers $n_{6,7,8,\ldots,}$ and $m_{1,\ldots, 8}\,$ that recover the superconformal index~\eqref{BA} at every possible value of $Y_1\,$; which corresponds to equating~$\mathcal{C}^\prime$ to the Bethe contour~$\mathcal{C}\,$.
This is an infinitely over-constrained system of equations. 
Generically, one would have expected not to have a solution to infinitely many constraints with simply $20$ integer variables to solve for. 
Fortunately, we have found that such a solution exists, and its existence is rather non-trivial. 
Indeed, if one simply ignores the contribution of some of the potentially contributing poles, the system of constraints suddenly becomes unsolvable. This is non-trivial evidence that equality~\eqref{BA} holds and that our proposed prescription to compute the residues of continuous Bethe roots works fine.

Each one of the individual contributions~\eqref{eq:NonTopContribRes} has tachyonic powers of negative order in small-$\widetilde{q}$ expansion. They come at order
\be
\frac{1}{\widetilde{q}^4}\,,\, \frac{1}{\widetilde{q}^3}\,,\, \frac{1}{\widetilde{q}^2}\,, \, \frac{1}{\widetilde{q}^1}
\ee
with coefficients that include non-rational functions of $Y_1\,$. The number of such independent functions and their corresponding numerical factors grow both very fast with the powers of $\widetilde{q}\,$. For instance, at power $\widetilde{q}^{0}$ there come already many independent functions of $Y_1$ with numerical coefficients of order~$10^{8}\,$. The desired linear combination of residues~\eqref{eq:NonTopContribRes} and~\eqref{ResiduesIrredTop} is such that both, the individual tachyonic contributions and the $10^8$ order coefficients, cancel among themselves to give
\begin{equation}
\mathcal{I} \,=\, 1\,+\,\ldots\,.
\end{equation}
These cancellations are rather non-trivial. They are convincing evidence in favour of the method here proposed.\footnote{
Ideally one would like to move on and reproduce the index $\mathcal{I}$ up to arbitrary powers in $\widetilde{q}$ starting from~\eqref{IndexToCompare}. In practice, however, with the symbolic computational power that we have currently at our disposal we decided to stop at testing the cancellations of tachyonic contributions and matching of the single-vacuum counting $1\,$.}

Next we move on to share details of the computation just described. Here we will only report intermediate results at leading tachyonic order $\frac{1}{\widetilde{q}^4}\,$. This is because the intermediate expressions involved at subleading order are huge and sharing them here will not add any value to the presentation.

\paragraph{Irreducible BC 1}
Next we cover all possible boundary conditions and find all possible contributing Bethe roots. For the irreducible boundary condition 1, the ansatz for the gauge rapidities takes the form
\be\label{eq:AnsatzIBC1}
\underline{z}=\underline{Z}:=\{Z_1,Z_2\}\,.
\ee
The integrand expanded around the ansatz can be recast in the following form
\be
{I}_{\textrm{B.A}} \,=\, \frac{K(\widetilde{q})}{U_1 U_2} \, {d Z_1}\,\wedge\, {dZ_2} = \frac{K(\widetilde{q})}{U_1 U_2}
\,\frac{d U_1 \wedge d U_2}{\det \left( \frac{\partial (U_1, U_2)}{\partial (Z_1, Z_2)} \right)}
\ee
where
\be
U_1= U(\underline{Z}): = (Z_1 Z_2 - 1), \quad U_2 = U_2(\underline{Z}):= (Z_2 - Z_1^2)
\ee
are $\widetilde{q}$-independent polynomials of $Z_{i}$'s and
\be
\begin{split}
K &\,:=\,\frac{ U_1 U_2\kappa I}{z_1(1-Q_1)z_2(1-Q_2)}\,=\, \sum_{n=0}^{\infty} {K^{(n)}}\widetilde{q}^{n} \\ &\,=\,- \frac{(-1 + Z_2)^2 (Z_2 - Z_1)^2 (-1 + Z_1)^2}{6 \left(Z_2^3 Z_1\right)} \,+\, \mathcal{O}(\widetilde{q})
\end{split}
\ee
is a holomorphic series whose coefficients have no poles in the interior of the integration contour $\mathcal{C}$ i.e.
\be
1\,\leq \,|z_i| \,\leq\, \frac{1}{|\widetilde{q}^3|}\,. 
\ee
All contributing poles come from the intersection
\be
U_1=U_2=0
\ee
which is the set of isolated points~\eqref{eq:N3DiscreteSols}
\be
\{\underline{Z}_{(0)}\}\,=\,\Bigl\{\{1,1\},\{-(-1)^{\frac{1}{3}},(-1)^{\frac{2}{3}}\},\{(-1)^{\frac{2}{3}},-(-1)^{\frac{1}{3}}\}\Bigr\}\,.
\ee
Evaluating the residue function
\be\label{ResFunCasek3}
\text{Res}[I_{\textrm{B.A}},{\underline{Z}_0}]\,=\,\frac{1}{2\pi\text{i}} \, \frac{K}{\text{det}_{i,j}\Bigl(\partial_{Z_i} U_{j}\Bigr)}\,{\Bigl|}_{\underline{Z}=\underline{Z}_{(0)}}
\ee
in the zero locus \eqref{eq:N3DiscreteSols} one obtains the residue contributions to the superconformal index coming from the first two irreducible solutions, $R_1$ and $R_2\,$, at any order in the small-$\widetilde{q}$ expansion
\be
\Bigl\{0,R_1\,:=\,\frac{3}{2}+\mathcal{O}(\widetilde{q}),R_2:=\frac{3}{2}\,+\,\mathcal{O}(\widetilde{q})\Bigr\}\,.
\ee
Note that
\be
R_1\,+\,R_2 \,=\, 3\,+\,\mathcal{O}(\widetilde{q})\,.
\ee
This is the contribution coming from residues of two irreducible topological roots.

\paragraph{Irreducible BC 2}
The ansatz takes the form
\be
\underline{z}= \{Z_1 \widetilde{q},\,Z_2 \widetilde{q}^2\}\,.
\ee
The integrand expanded around the ansatz can be recast in the following form
\be
{I}_{\textrm{B.A}} \,=\, \frac{K(\widetilde{q})}{U_1 U_2 U_3^2} \, {d Z_1}\,\wedge\, {dZ_2} = \frac{K(\widetilde{q})}{U_1 U_2 U_3^2}
\,\frac{d U_i \wedge d U_j}{\text{det}_{i,j}\Bigl(\partial_{Z_i} U_{j}\Bigr)}
\ee
where
\be\label{eq:U3BC2}
U_3 := Y_1 Y_2 Z_1^2+Y_1 Y_2 Z_2^2 Z_1-Z_2 \left(Y_2 Y_1^2 Z_1+Y_2 Y_1 \left(Y_2
   Z_1-1\right)+Z_1\right)
\ee
is a $\widetilde{q}$-independent polynomial of $Z_{i}$'s and
\be
\begin{split}
K &\,:=\,\frac{ U_1 U_2 U_3^2\kappa I}{z_1(1-Q_1)z_2(1-Q_2)}\,=\,\sum_{n=0}^{\infty} \frac{K^{(n)}}{\bigl(U_3\bigr)^n}\widetilde{q}^{\,n-4}  \\
   &\,=\,-\frac{Y_1 Y_2 Z_1^3 Z_2 \left(Z_2-Y_1\right) \left(Z_2-Y_2\right) \left(Y_1 Y_2
   Z_2-1\right)}{6 \widetilde{q}^4} \,+\, \mathcal{O}(\frac{1}{\widetilde{q}^3})
   \end{split}
\ee
where $K^{(n)}$ are holomorphic functions with no poles in the interior of the integration contour $\mathcal{C}\,$ and independent of $\widetilde{q}\,$.
All contributing poles come from the three possible intersections~$k=1,2,3$
\be
\{\underline{Z}_{(0,k)}\}: U_i=U_j=0\,, \qquad i\neq j\neq k\,,\qquad i,j,k =1,2,3.
\ee
\underline{The simplest case is the intersection $k=3\,$:} 
 this is the intersection
\be
U_1=U_2=0
\ee
which gives the set of isolated points
\be
\{\underline{Z}_{(0,3)}\}\,=\,\Bigl\{\{1,1\},\{-(-1)^{\frac{1}{3}},(-1)^{\frac{2}{3}}\},\{(-1)^{\frac{2}{3}},-(-1)^{\frac{1}{3}}\}\Bigr\}\,.
\ee
Evaluating the residue function
\be
\text{Res}[I_{\textrm{B.A}},{\underline{Z}_{(0,3)}}]\,= \, \frac{K}{U_3^2\,\text{det}_{i,j}\Bigl(\partial_{Z_i} U_{j}\Bigr)}\,{\Bigl|}_{\underline{Z}=\underline{Z}_{(0,3)}}
\ee
in these poles one obtains their residue contributions to the superconformal index, $R_3$, $R_4$, and $R_5$ at leading order in the small-$\widetilde{q}$ expansion
\be
\begin{split}
\left\{R_3\,:=\,-\frac{Y_1^2 \left(Y_1+1\right)}{18 \widetilde{q}^4 \left(Y_1-1\right) \left(2 Y_1+1\right)^2}+\mathcal{O}\left(\frac{1}{\widetilde{q}^3}\right),\right.\\ \left.R_4\,:=\,-\frac{\left((-1)^{2/3}-Y_1\right)^2 Y_1^2 \left((-1)^{2/3} Y_1^2-1\right)}{18 \widetilde{q}^4\left(2
   Y_1^3-3 (-1)^{2/3} Y_1^2+1\right)^2}+\mathcal{O}\left(\frac{1}{\widetilde{q}^3}\right),\right. \\ \left. R_5 \,:=\,\frac{Y_1^2 \left(Y_1+(-1)^{1/3}\right)^2 \left((-1)^{1/3} Y_1^2+1\right)}{18
  \widetilde{q}^4 \left(2 Y_1^3+3 (-1)^{1/3} Y_1^2+1\right)^2}+\mathcal{O}\left(\frac{1}{\widetilde{q}^3}\right)\right\}
   \end{split}
\ee

Note that
\be
R_3\,+\,R_4\,+\,R_5\,=\, -\frac{Y_1^4 \left(2 Y_1^3+1\right)}{\widetilde{q}^4 \left(Y_1^3-1\right) \left(8
   Y_1^3+1\right){}^2}+\mathcal{O}\left(\frac{1}{\widetilde{q}^3}\right)
\ee
This is the contribution coming from residues of the another three irreducible topological roots.

\underline{Intersection $k=2\,$:}
This is the intersection
\be
U_1=U_3=0
\ee
which gives the set of isolated points
\be
\begin{split}
\{\underline{Z}_{(0,2)}\}\,=\,\left\{\left\{\frac{1}{Y_1}, Y_1\right\},\left\{\frac{-\sqrt{8
   Y_1^3+1}-1}{2 Y_1}, \frac{1-\sqrt{8 Y_1^3+1}}{4 Y_1^2}\right\},\right. \\ \left.\left\{
   \frac{\sqrt{8 Y_1^3+1}-1}{2 Y_1}, \frac{\sqrt{8 Y_1^3+1}+1}{4
   Y_1^2}\right\}\right\}\,.
\end{split}
\ee
Their residues can be evaluated as follows
\be
\begin{split}
\text{Res}[I_{\textrm{B.A}},{\underline{Z}_{(0,2)}}]&\,=\, { \frac{\partial \underline{Z} }{\partial  U_3} \cdot \frac{\partial}{\partial_{\underline{Z}}} \Biggl(\frac{\mathfrak{m}_{\underline{Z}}}{\det \left( \frac{\partial (U_1, U_3)}{\partial (Z_1, Z_2)} \right)}\,\frac{K}{U_2}\Biggr)}\,{\Biggl|}_{\underline{Z}=\underline{Z}_{(0,2)}}\\ & \,=\, {\sum_{i=1}^{2}\left( \frac{\partial (U_1, U_3)}{\partial (Z_1, Z_2)} \right)^{-1}_{i2} \frac{\partial}{\partial_{Z_i}}  \Biggl(\frac{\mathfrak{m}_{\underline{Z}}}{\det \left( \frac{\partial (U_1, U_3)}{\partial (Z_1, Z_2)} \right)}\,\frac{K}{U_2}\Biggr)}\,{\Biggl|}_{\underline{Z}=\underline{Z}_{(0,2)}}
\end{split}
\ee
where~$\mathfrak{m}_{\underline{Z}}\,\in\,\mathbb{Z}$ and
\be
\begin{split}
\left( \frac{\partial (U_1, U_3)}{\partial (Z_1, Z_2)} \right)&\,: =\, \begin{pmatrix} \partial_{Z_1} U_1 & \partial_{Z_2} U_1 \\ \partial_{Z_1} U_3 & \partial_{Z_2} U_3 \end{pmatrix}\,, \qquad  \left( \frac{\partial (U_1, U_3)}{\partial (Z_1, Z_2)} \right)^{-1}\,:=\, \begin{pmatrix} \partial_{U_1} Z_{1} & \partial_{U_3} Z_1 \\ \partial_{U_1} Z_2 & \partial_{U_3} Z_2 \end{pmatrix}
\,.
\end{split}
\ee
The multivariate residues of these poles depends on how $\mathcal{C}$ surrounds them. This is the meaning of the integer ambiguities~$\mathfrak{m}_{\underline{Z}}\,$, they precise how $\mathcal{C}$ surrounds the corresponding pole.

Using the formula above we obtain
\be
\begin{aligned}
\left\{0\,,\,R_6\, :=\, n_6\, \frac{Y_1^4-Y_1^7 \left(\sqrt{8 Y_1^3+1}-2\right)}{2 \widetilde{q}^4 \left(Y_1^3-1\right)
   \left(8 Y_1^3+1\right){}^2} + \mathcal{O}\left(\frac{1}{\widetilde{q}^3}\right)\,, \right. \\
\left. R_7\,:=\, n_7\,\frac{\left(\sqrt{8 Y_1^3+1}+2\right) Y_1^7+Y_1^4}{2 \widetilde{q}^4 \left(Y_1^3-1\right)
   \left(8 Y_1^3+1\right){}^2} + \mathcal{O}\left(\frac{1}{\widetilde{q}^3}\right)\right\}
\end{aligned}
\ee
where
\be
n_{6}\,, \, n_7\,\in\,\mathbb{Z}
\ee
correspond to the ambiguity in how $\mathcal{C}^\prime$ surrounds these poles.

\underline{Intersection $k=1\,$:} This is the intersection
\be
U_2=U_3=0
\ee
which gives the set of isolated points
\be
\begin{split}
\{\underline{Z}_{(0,1)}\}\,=\,\Biggl\{\left\{\frac{1}{Y_1},\frac{1}{Y_1^2}\right\},&\left\{
   \frac{-\sqrt{8 Y_1^3+1}-1}{2 Y_1},\frac{1}{2} \left(4 Y_1+\frac{\sqrt{8
   Y_1^3+1}}{Y_1^2}+\frac{1}{Y_1^2}\right)\right\}, \\  &\left\{ \frac{\sqrt{8
   Y_1^3+1}-1}{2 Y_1}, \frac{1}{2} \left(4 Y_1-\frac{\sqrt{8
   Y_1^3+1}}{Y_1^2}+\frac{1}{Y_1^2}\right)\right\}\Biggr\}
   \end{split}
\ee
Their residues can be evaluated as follows
\be
\text{Res}[I_{\textrm{B.A}},{\underline{Z}_{(0,1)}}]\,=\,{ \frac{\partial \underline{Z} }{\partial  U_3} \cdot \frac{\partial}{\partial_{\underline{Z}}} \Biggl(\frac{\mathfrak{m}_{\underline{Z}}}{\det \left( \frac{\partial (U_2, U_3)}{\partial (Z_1, Z_2)} \right)}\,\frac{K}{U_1}\Biggr)}\,{\Biggl|}_{\underline{Z}=\underline{Z}_{(0,1)}}\,.
\ee
They are
\be
\begin{split}
\Biggl\{R_8&\,:=\,- \,\frac{n_8 Y_1}{12 \widetilde{q}^4 \left(Y_1^3-1\right)}+\mathcal{O}\left(\frac{1}{\widetilde{q}^3}\right),\\R_9&\,:=\,\frac{n_9 Y_1 \left(4 \left(-5 \left(\sqrt{8 Y_1^3+1}-2\right) Y_1^3+\sqrt{8
   Y_1^3+1}+1\right) Y_1^3+\sqrt{8 Y_1^3+1}+1\right)}{24 \widetilde{q}^4
   \left(Y_1^3-1\right) \left(8 Y_1^3+1\right){}^2}\\&+\mathcal{O}\left(\frac{1}{\widetilde{q}^3}\right),\\R_{10}&\,:=\,n_{10}\,\frac{20 \left(\sqrt{8 Y_1^3+1}+2\right) Y_1^7-4 \left(\sqrt{8 Y_1^3+1}-1\right)
   Y_1^4-\sqrt{8 Y_1^3+1} Y_1+Y_1}{24 \widetilde{q}^4 \left(Y_1^3-1\right) \left(8
   Y_1^3+1\right){}^2}\\&+\mathcal{O}\left(\frac{1}{\widetilde{q}^3}\right)\Biggr\}
   \end{split}
\ee
where
\be
n_{9}\,,\,n_{10}\,\in \, \mathbb{Z}\,
\ee
correspond to the ambiguity in how $\mathcal{C}^\prime$ surrounds these poles.

\paragraph{Irreducible BC 3}
The ansatz takes the form
\be
\underline{z}= \{Z_1 \widetilde{q}^2,\,Z_2 \widetilde{q}\}
\ee
The integrand expanded around the ansatz can be recast in the following form
\be\label{eq:BABC3}
{I}_{\textrm{B.A}} \,=\, \frac{K(\widetilde{q})}{U_1 U_2 U_3^2 U_4 U_5^2} \, {d Z_1}\,\wedge\, {dZ_2} = \frac{K(\widetilde{q})}{U_1 U_2 U_3^2 U_4 U_5^2}
\,\frac{d U_i \wedge d U_j}{\text{det}_{i,j}\Bigl(\partial_{Z_i} U_{j}\Bigr)}
\ee
with $\{U_i,U_j\}\,=\,\{U_1,U_2\}\,,\text{or } \{U_2,U_3\}\,,\text{or } \{U_1,U_3\}\,$. There will be intersections involving $U_4$ and $U_5$ that will be conveniently analyzed later. $U_1$ and $U_2$ are the same as before, instead
\be\label{eq:U345}
\begin{split}
U_3 &\,:=\, Y_1^2 Z_2^2+Y_1^2 Z_1+Z_1 Z_2 \left(Y_1^2 Z_1-2 Y_1^3-1\right)\,, \\
U_4&\,:=\,\left(Z_2-Y_1^2\right)\,,\quad
U_5\,:=\,\left(Y_1 Z_2-1\right)\,,
\end{split}
\ee
and
\be
\begin{split}
K&\,:=\,\frac{ U_1 U_2 U_3^2 U_4 U_5^2\kappa I}{z_1(1-Q_1)z_2(1-Q_2)}\,=\,\sum_{n=0}^{\infty} \frac{K^{(n)}}{\bigl(U_3\bigr)^n}\widetilde{q}^{\,n-4}\\ &\,=\,\frac{Y_1^2 Z_2^3 \left(Y_1-Z_1\right){}^2 \left(Y_1^2 Z_1-1\right) \left(Z_2-Y_1
   Z_1\right){}^2 \left(Y_1^2 Z_2-Z_1\right)}{6 \widetilde{q}^4 Z_1 
}\,+\,\mathcal{O}\bigl(\frac{1}{\widetilde{q}^3}\bigr)
\end{split}
\ee
where the $K^{(n)}$ are holomorphic functions of $\underline{Z}$, and independent on $\widetilde{q}\,$, with no poles in the interior of the integration contour $\mathcal{C}\,$.
We start computing residues coming from the three possible intersections~$k=1,2,3$
\be
\{\underline{Z}_{(0,k)}\}: U_i=U_j=0\,, \qquad i\neq j\neq k\,,\qquad i,j,k =1,2,3.
\ee
At last we move on to compute residues involving triple intersections with either $U_4$ or $U_5$, which are simpler.

\underline{Intersection $k=3\,$:}

\be
U_1=U_2=0
\ee
which gives the very same set of isolated points as before
\be
\{\underline{Z}_{(0,3)}\}\,=\,\Bigl\{\{1,1\},\{-(-1)^{\frac{1}{3}},(-1)^{\frac{2}{3}}\},\{(-1)^{\frac{2}{3}},-(-1)^{\frac{1}{3}}\}\Bigr\}\,.
\ee
Evaluating the residue function formula~\eqref{ResFunCasek3} with the new definitions of $U_3$ and $K$ in these poles, one obtains their residue contributions to the superconformal index, $R_{11}$, $R_{12}$, and $R_{13}$ at any order in the small-$\widetilde{q}$ expansion
\be
\begin{split}
\left\{R_{11}\,:=\,\frac{\left(1-Y_1\right)^2 Y_1^2 \left(Y_1^2-1\right)^2}{18 \widetilde{q}^4
   \left(1-Y_1^2\right) \left(\left(1-2 Y_1\right) Y_1^2+2 Y_1^2-1\right)^2}+\mathcal{O}\left(\frac{1}{\widetilde{q}^3}\right),\right.\\ \left.R_{12}\,:=\, \frac{\left(Y_1+(-1)^{\frac{1}{3}}\right)^3 \left((-1)^{\frac{1}{3}} Y_1^3+Y_1\right)^2}{18
   \widetilde{q}^4 \left((-1)^{\frac{1}{3}}-Y_1\right) \left((-1)^{\frac{2}{3}} Y_1-1\right)^2 \left(2
   Y_1^3+3 (-1)^{\frac{1}{3}} Y_1^2+1\right)^2}+\mathcal{O}\left(\frac{1}{\widetilde{q}^3}\right),\right. \\ \left. R_{13} \,:=\,-\frac{\left((-1)^{\frac{2}{3}}-Y_1\right)^2 Y_1^2 \left((-1)^{\frac{2}{3}} Y_1^2-1\right)}{18
   \widetilde{q}^4 \left(2 Y_1^3-3 (-1)^{\frac{2}{3}} Y_1^2+1\right)^2}+\mathcal{O}\left(\frac{1}{\widetilde{q}^3}\right)\right\}
   \end{split}\,.
\ee
Note that
\be
R_{11}\,+\,R_{12}\,+\, R_{13}\,:=\,-\frac{Y_1^4 \left(2 Y_1^3+1\right)}{\widetilde{q}^4 \left(Y_1^3-1\right) \left(8
   Y_1^3+1\right){}^2}+\mathcal{O}\left(\frac{1}{\widetilde{q}^3}\right)\,.
\ee
This is the contribution coming from residues of the last three irreducible topological roots.

\underline{Intersection $k=2\,$:}
This is the intersection
\be
U_1=U_3=0
\ee
which gives the set of isolated points
\be
\begin{split}
\left\{\underline{Z}_{(0,2)}\right\}\,=\,\Biggl\{\left\{Y_1,\frac{1}{Y_1}\right\},&\left\{-\frac{\sqrt{8
   Y_1^3+1}-1}{4 Y_1^2},-\frac{\sqrt{8 Y_1^3+1}+1}{2 Y_1}\right\},\\&\left\{\frac{\sqrt{8 Y_1^3+1}+1}{4 Y_1^2},\frac{\sqrt{8 Y_1^3+1}-1}{2
   Y_1}\right\}\Biggr\}\,.
\end{split}
\ee
The multivariate residues of these poles depends on how $\mathcal{C}$ surrounds them (e.g. on the order of integration). They can be
\be
\begin{split}
\Biggl\{\star\,,&\,\\
R_{14}&\,=\,\frac{n_{14}Y_1 \left(20 \left(\sqrt{8 Y_1^3+1}-2\right) Y_1^6-4 \left(\sqrt{8
   Y_1^3+1}+1\right) Y_1^3-\sqrt{8 Y_1^3+1}-1\right)}{24 \widetilde{q}^4
   \left(Y_1^3-1\right) \left(8 Y_1^3+1\right){}^2}\\&+\mathcal{O}\left(\frac{1}{\widetilde{q}^3}\right),\\ R_{15}&\,=\,\frac{n_{15}Y_1 \left(-20 \left(\sqrt{8 Y_1^3+1}+2\right) Y_1^6+4 \left(\sqrt{8
   Y_1^3+1}-1\right) Y_1^3+\sqrt{8 Y_1^3+1}-1\right)}{24 \widetilde{q}^4
   \left(Y_1^3-1\right) \left(8 Y_1^3+1\right){}^2}\\&+\mathcal{O}\left(\frac{1}{\widetilde{q}^3}\right)\Biggr\}\,,
\end{split}
\ee
where
\be
n_{14}\,, \, n_{15}\,\in\, \mathbb{Z}\,,
\ee
classify the aforementioned ambiguity.

{The $\star$ denotes the fact that this pole comes really from a triple intersection}
\be
U_2=U_3=U_5=0\,.
\ee
It is also a zero of the numerator. A computation shows that the residue of this pole vanishes.

\underline{Intersection $k=1\,$:} this is the intersection
\be
U_2=U_3=0
\ee
which gives the set of isolated points
\be
\begin{split}
\left\{\underline{Z}_{(0,1)}\right\}\,=\,\Biggl\{\left\{Y_1,{Y_1}^2\right\},&\left\{-\frac{\sqrt{8 Y_1^3+1}-1}{4 Y_1^2},\frac{4 Y_1^3-\sqrt{8 Y_1^3+1}+1}{8
   Y_1^4}\right\},\\&\left\{\frac{\sqrt{8 Y_1^3+1}+1}{4 Y_1^2},\frac{4 Y_1^3+\sqrt{8 Y_1^3+1}+1}{8
   Y_1^4}\right\}\Biggr\}\,.
\end{split}
\ee
Their residues are
\be
\begin{split}
\Biggl\{\,\star\,,&\,\\
R_{16}&\,=\,\frac{n_{16}Y_1 \left(44 \left(\sqrt{8 Y_1^3+1}-2\right) Y_1^6-4 \left(\sqrt{8
   Y_1^3+1}+7\right) Y_1^3-\sqrt{8 Y_1^3+1}-1\right)}{24 \widetilde{q}^4
   \left(Y_1^3-1\right) \left(8 Y_1^3+1\right){}^2}\\ &+\mathcal{O}\left(\frac{1}{\widetilde{q}^3}\right),\\ R_{17}&\,=\,\frac{n_{17}Y_1 \left(-44 \left(\sqrt{8 Y_1^3+1}+2\right) Y_1^6+4 \left(\sqrt{8
   Y_1^3+1}-7\right) Y_1^3+\sqrt{8 Y_1^3+1}-1\right)}{24 \widetilde{q}^4
   \left(Y_1^3-1\right) \left(8 Y_1^3+1\right){}^2}\\ &+\mathcal{O}\left(\frac{1}{\widetilde{q}^3}\right)\Biggr\}
\end{split}
\ee
where
\be
n_{16}\,,\, n_{17}\,\in \,\mathbb{Z}\,
\ee
classify the aforementioned ambiguity.

{The $\star$ denotes the fact that this pole comes really from a triple intersection}
\be
U_2=U_3=U_4=0\,.
\ee
A computation below will show that the residue of this pole may be non-vanishing.

\paragraph{Poles from triple intersections (BC 3)}

There are other 4 contributing poles that come from triple intersections including one of the two remaining denominator-factors~$U_4$ and~$U_5$
\be
(Z_{2}\,-\,Y_1^2)\,, \qquad ({Y_1} Z_2\,-\,1)\,.
\ee
They are
\be
\{\underline{Z}_{0}\} =\Bigl\{\bigl\{\pm Y_1,Y_1^2\bigr\},\bigl\{\pm \frac{1}{\sqrt{Y_1}},\frac{1}{Y_1}\bigr\}\Bigr\}\,.
\ee
Their residues are
\be
\begin{split}
\Bigl\{ R_{18}&\,=\,n_{18}\frac{-Y_1}{12 \widetilde{q}^4 \left(Y_1^3-1\right)}\,+\, \mathcal{O}\Bigl(\frac{1}{\widetilde{q}^3}\Bigr)\,, \, R_{19}\,=\,n_{19}\frac{-Y_1^4-Y_1}{12 \widetilde{q}^4 \left(Y_1^3-1\right){}^2}\,+\, \mathcal{O}\Bigl(\frac{1}{\widetilde{q}^3}\Bigr) \,,\\R_{20}&\,=\, n_{20}\frac{Y_1^4+3 Y_1^{5/2}}{24 \widetilde{q}^4 \left(Y_1^{3/2}-1\right)
   \left(Y_1^{3/2}+1\right){}^2} \,+\, \mathcal{O}\Bigl(\frac{1}{\widetilde{q}^3}\Bigr)\,, \\ R_{21}&\,=\,n_{20}\frac{-Y_1^4+3 Y_1^{5/2}}{24 \widetilde{q}^4 \left(Y_1^{3/2}-1\right){}^2
   \left(Y_1^{3/2}+1\right)} \,+\, \mathcal{O}\Bigl(\frac{1}{\widetilde{q}^3}\Bigr)\, \Bigr\}
\end{split}
\ee
where
\be
n_{18}\,,\, n_{19}\,,\, n_{20} \,\in\, \mathbb{Z}\,,
\ee
classify the aforementioned ambiguity. Note that we have chosen~$n_{21}=n_{20}$ because $R_{20}$ and $R_{21}$ are the only two potential residues with explicit odd powers in $Y^{\frac{1}{2}}\,$. These powers can not be there in the index so imposing this condition implies $n_{21}\,=\,n_{20}\,$.

\subsubsection{Reducible boundary conditions}

Now we follow the same steps as before but for the reducible boundary conditions. This time the ansatze are (within the annulus~\eqref{eq:annulus}):

\begin{itemize}

\item
\textbf{Reducible BC 1}\qquad 
$\underline{z}= \{Z_1,\,Z_2 \widetilde{q}^{\frac{3}{2}}\}\,\sim\, \{Z_1,\,\frac{Z_2}{\widetilde{q}^{\frac{3}{2}}}\}$\,.

\item \textbf{Reducible BC 2}$\qquad \underline{z}\,=\, \{Z_1 \widetilde{q}^{\frac{3}{2}},\,Z_2\}\sim \{\frac{Z_1}{\widetilde{q}^{\frac{3}{2}}},Z_2\}$\,.

\item \textbf{Reducible BC 3}\qquad $
\underline{z}= \{Z_1 \widetilde{q}^{\frac{3}{2}},\,Z_2 \widetilde{q}^{\frac{3}{2}}\}\,\sim\,\{\frac{Z_1}{\widetilde{q}^{\frac{3}{2}}},\,
\frac{Z_2}{\widetilde{q}^{\frac{3}{2}}}\}.$

\end{itemize}
All in all, we find 8 possibly non-vanishing residues coming from these boundary conditions, which we denote as
\be
\begin{split}
\Biggl\{R_{21}&\,=\,-\frac{2 m_1}{3 \widetilde{q}^3}\,+\, \mathcal{O}\Bigl(\frac{1}{\widetilde{q}^{\frac{5}{2}}}\Bigr)  \,,\, R_{22}\,=\, \frac{2m_2}{3 \widetilde{q}}\,+\, \mathcal{O}\Bigl(\frac{1}{\widetilde{q}^{\frac{5}{2}}}\Bigr)\,,\,R_{23}\,=\,-\frac{2 m_3}{3 \widetilde{q}^3}+\, \mathcal{O}\Bigl(\frac{1}{\widetilde{q}^{\frac{5}{2}}}\Bigr)\,,\\ \quad \,R_{24}&\,=\,\frac{2 m_4}{3 \widetilde{q}^3}\,+\, \mathcal{O}\Bigl(\frac{1}{\widetilde{q}^{\frac{5}{2}}}\Bigr)\,,\, R_{25}\,=\,\frac{m_5}{12 \widetilde{q}^3}\,+\, \mathcal{O}\Bigl(\frac{1}{\widetilde{q}^{\frac{5}{2}}}\Bigr)\,, \, R_{26}\,=\,\frac{-m_6}{12 \widetilde{q}^3}\,+\, \mathcal{O}\Bigl(\frac{1}{\widetilde{q}^{\frac{5}{2}}}\Bigr)\,,\\ R_{27}&\,=\,0\,+\, \mathcal{O}\Bigl(\frac{1}{\widetilde{q}^{\frac{5}{2}}}\Bigr)\,,\, R_{28}\,=\,\frac{m_8}{12 \widetilde{q}^3}\,+\, \mathcal{O}\Bigl(\frac{1}{\widetilde{q}^{\frac{5}{2}}}\Bigr)\,\Biggr\}\,,
\end{split}
\ee
with
\be
m_{1,\ldots,8}\,\in\,\mathbb{Z}
\ee
classifying the aforementioned ambiguity.

\subsection{Cancelling tachyonic contributions}

Collecting residues of all poles selected by irreducible boundary conditions one obtains
\be
R_{irred}\,=\,\sum^{20}_{n=1}R_n
\ee
Collecting also all possible poles selected by reducible boundary conditions one obtains
\be
\begin{split}
R_{red}&\,=\,\sum^{28}_{n=21}R_n\,.
\end{split}
\ee
Then imposing
\be\label{eq:ResidueMatching}
R_{irred}\,+\, R_{red}\,=\, 1\,+\,\mathcal{O}(\widetilde{q})\,,
\ee
we find that the initial 17 free parameters must obey 7 constraints. This is, there is a 10-parametric family of residue selections guarantying~\eqref{eq:ResidueMatching}  
\be\label{eq:SolCancTachyons}
\begin{split}
& n_6 = c_2 + 2c_4, \, n_7 = c_2 + 2c_5, \, n_8 = c_2, \, n_9 = c_6, \, n_{10} = c_7, \\
   & n_{14} = 1 - c_2 - c_4 + c_6, \, n_{15} = 1 - c_2 - c_5 + c_7, \\
   & n_{16} = -1 + c_4, \, n_{17} = -1 + c_5\,,\\
   &  n_{18} = 0, \, n_{19} = 0, \, n_{20} = 0, \\
   & m_1 = c_8, \, m_2 = c_9, \, m_3 = c_1 + c_3 + 4c_8 + 5c_{10}, \\
   & m_4 = c_1 + c_2 + c_3 + 5c_8 - c_9 + 5c_{10}, \, m_5 = c_3, \\
   & m_6 = 2 + c_1 + 2c_2 + c_3, \, m_7 = c_1 + c_3 + c_8 + c_{10}, \, m_8 = c_1\,.
\end{split}
\ee
The arbitrary integer parameters are $c_{1,\ldots, 10}\,$.

To reach~\eqref{eq:SolCancTachyons} all tachyonic contributions of order
\be
\widetilde{q}^{-4}\,,\, \widetilde{q}^{-3}\,,\, \widetilde{q}^{-2}\,, \,\widetilde{q}^{-1}
\ee
coming from topological poles has been cancelled by the contribution coming from non-topological roots. Note that these cancellations happened at all possible values of $Y_1\,$. So in a sense, the 7 relations imposed on the initial 17 integer variables to reach the 10-dimensional parameterization~\eqref{eq:SolCancTachyons} have solved infinitely many constraints. 

\begin{table}[h]
    \centering
    \begin{tabular}{|c|c|}
        \hline
        Boundary conditions & $h(\underline{z})$ \\ \hline
       \textbf{Irreducible BC2}\, $\longrightarrow$\, $\gamma_1$  & $U_{3, \text{ in eq.~\eqref{eq:U3BC2}}}\,+\,\ldots$  \\ \hline
        \textbf{Irreducible BC3}\, $\longrightarrow$\, $\gamma_2$ and $\gamma_3$  & $U_{3 \text{ or }5, \text{ in eq.~\eqref{eq:U345}}}\,+\,\ldots$   \\ \hline
        \textbf{Reducible BC2} \, $\longrightarrow$\, $\gamma_4$ & $\,\,\,\,\,\,\,\,Z_{1}^2\,+\, Z_2\,\,+\,\ldots\,=\, z_1^2\,+\, z_2\,{q}^{\frac{1}{2}} \,+\,\ldots$  \\ \hline 
           \textbf{Reducible BC3} \, $\longrightarrow$\, $\gamma_5$      &           $Z_{2} Z_1+1\,+\,\ldots\,=\,z_2 z_1\, {q}^{\frac{1}{2}} \,+\,\ldots $ \\\hline
    \end{tabular}
    \caption{Independent choices of the functions $h$'s defined in~\eqref{eq:Defh}. Each choice corresponds to an independent continuous solution~$\gamma\,$. The $\ldots$ denote subleading corrections in the corresponding expansion around ${q}=0\,$. It is also possible that some of these local choices of $h$ come the same global function $h$ expanded around different boundary conditions.}
    \label{tab:two_columns}
\end{table}

\section{Summary and discussion}

Given a superconformal index $\mathcal{I}\,$, we have shown that there are infinitely many possible Bethe expansions associated to it, with some of them being more convenient to use than others. 
All of them compute $\mathcal{I}$ as a finite sum over solutions to Bethe equations.
Our formalism applies to general superconformal theories, and maybe even beyond.\footnote{
Technically, it applies to any integral of the form~\eqref{eq:TheIntegral} with a meromorphic integrand~$I$ of the form~\eqref{eq:IntegrandProd} that obeys properties~\ref{property1},~\ref{property2}, and~\ref{property3}.} 

\medskip

Using this formalism, we have proposed solutions to two main problems:
\begin{itemize}
\item[1.] Find all the Bethe solutions that (potentially) contribute to the superconformal index.
\item[2.] 
Evaluate the contribution to the index from the continuous solutions.
\end{itemize}

We have already given a summary of general results in Section~\ref{sec1}.
In this final section we focus on summarizing the salient points for the concrete example of $U(N)$ $\mathcal{N}=4$ SYM with $p=q\,$.

First, let us summarize our results for problem (1).

Topological solutions are distinguished by their behavior under cyclic symmetries. For example, the irreducible topological solutions are representations of $\mathbb{Z}_N\,$, the center symmetry within the Weyl symmetry group of~$U(N)$ $\mathcal{N}=4$ SYM. On the other hand, the reducible topological solutions are made of representations of smaller cyclic groups $\mathbb{Z}_{N'}$ with $N'<N$.

\smallskip

Non-topological solutions can not be determined using cyclic symmetries. To compute these we use trigonometric-like expansions near $q=0$ under which the Bethe equations simplify. 
In fact, these expansions systematically determine all solutions, both topological and non-topological.
These expansions are fixed by certain boundary conditions which are determined by the set of all possible asymptotic expansions of both topological and non-topological solutions near $q=0\,$. We claim that topological and non-topological solutions share the same set of leading asymptotic behaviors near $q=0$.\footnote{Initially, we were led to explore this possibility because tachyonic contributions coming from topological solutions in the expansion around $q=0\,$, needed to be cancelled by tachyonic contributions coming from non-topological solutions in the very same expansion. }
Thus, one can first determine all the topological solutions using cyclic symmetries, then use them to fix all the trigonometric-like expansions, and finally use these expansions to obtain all the non-topological solutions in perturbations around $q=0\,$.

Under these trigonometric-like expansions, the Bethe equations become rational in fugacities, and the topological solutions are determined by $N$ zeroes of a finite number of order-$N$ polynomials in a single complex variable. 
The Galois group of one generic such polynomial is the symmetric group $S_N\,$. 
The breaking of such Galois group indicates the emergence of a spontaneously broken discrete symmetry that exchanges the corresponding $N$ roots among themselves,\footnote{
This is $\mathbb{Z}_N$ one-form center symmetry identified in~\cite{Cabo-Bizet:2021jar}.} which is the $\mathbb{Z}_{N}\,$ group, and the corresponding solutions are the irreducible topological solutions.

For contributing non-topological solutions, the Galois group of the corresponding order-$N$ polynomial remains unbroken. 
In particular, the $\mathbb{Z}_{N}$ symmetry is explicitly broken by these solutions. 
This can be immediately seen from their positions: in contrast to the topological solutions, the positions of non-topological ones depend on the flavour rapidities, e.g., on $Y_{1}\,$.

For generic superconformal indices, we expect the relevant trigonometric-like expansions to be fixed by the balancing condition~\eqref{eq:BalancingCondition}. 
However, further study is needed to put this conclusion on firm ground.

\bigskip

Regarding problem (2), we have shown that the contribution of non-topological solutions to the index localizes to isolated points in $\gamma\,$.
The details of this localization process have been illustrated in the concrete example of $N=3\,$.  

First of all, continuous solutions $\gamma$ intersect with the reducible topological solutions. 
We have shown that the reducible 
topological solutions do not contribute to the index.
The reason is that the contribution of $\gamma$'s gets localized to isolated points for which there is no cyclic-symmetry enhancement, whereas the reducible topological solutions still obey some cyclic symmetry.

Then for $N=3$, we have checked that the tachyonic contributions coming from the irreducible topological solutions are cancelled by tachyonic contributions from the non-topological solutions, in a rather non-trivial way that involves cancellations among several large rational coefficients of order $10^{8}\,$.
We expect these cancellations to continue to hold for generic values of $N\,$. 
This is in a sense predicted by the generality of the rather simple residue identity~\eqref{eq:ResFormula}, which relies solely on the meromorphicity of $I\,$ and the $t=q$-periodicity property~\eqref{eq:PeriodicityQOper}.

\bigskip

Finally, we mention some interesting future directions:
\begin{itemize}
\item It would be interesting to apply our formalism to study systems beyond the 4D superconformal indices. For example, our methods can be applied to study four-dimensional superconformal indices in the presence of superconformal defects~\cite{Chen:2023lzq,Cabo-Bizet:2023ejm,Amariti:2024bsr}. They can be also applied to study four-dimensional topologically twisted indices~\cite{Hong:2018viz}.

\item Instead of the approach we have followed in this paper, Bethe expansions can be also derived using non-abelian equivariant localization formulae. The details of such derivation will be reported in version 2 of~\cite{Cabo-Bizet:2020ewf}. 

\item For $\mathcal{N}=4$ SYM, it would be nice to reproduce the index up to higher orders in powers of $q$ and for higher values of~$N$.

In particular, we would like to determine the large-$N$ behavior of the continuous solutions and to understand their  gravitational/stringy bulk duals.

\item Also for $\mathcal{N}=4$ SYM, we have presented a simpler Bethe expansion for generic cases $p\neq q$,~\eqref{eq:QPQOperator}. We expect that this expansion, together with the proposed methods, will lead to the resolution of the puzzles recently raised in~\cite{Aharony:2024ntg}.

\end{itemize}

\section*{Acknowledgments}

We are grateful for the support and hospitality from the Issac Newton Institute for Mathematical Sciences at Cambridge University (during the Program ``Black holes: bridges between number theory and holographic quantum information" with EPSRC Grant ER/R014604/1), where this project was started. 
ACB is supported by the INFN grant GAST, and is grateful for the support and hospitality from the INFN Galileo Galilei Institute for Theoretical Physics (during the programme “Resurgence and Modularity in QFT and String Theory”), where part of this work was carried out.
WL is supported by NSFC  No.\ 12275334, No.\ 11947302 and is grateful for the hospitality of the Kavli Institute for Theoretical Physics at Santa Barbara and 
SUSTech Center for Mathematics, where part of this work was carried out. 

\appendix

\section{Elliptic functions}
The~$q$-Pochammer symbol~$(\zeta;q)\equiv (\zeta;q)_{\infty}$ has the following product representation~\cite{1999math......7061F}
\begin{equation}
(\zeta;q)\= \prod_{j=0}^{\infty}(1\,-\,q^{j}\,\zeta)\,.
\end{equation}
The quasi-elliptic function $\theta_0$ has the following product representation
\begin{equation}\label{ThetaDef}
\theta_0(\z;q) \,=\,  (1-\z) \prod_{j=1}^\infty (1-q^j \z) \, (1-q^j \z^{-1}) \,.
\end{equation}
The following properties were used in the main body of the paper
\be\label{eq:QuasiPThetas}
\theta_0(q\zeta;q)\,=\,-\frac{\theta_0(\zeta;q)}{\zeta}\,, \qquad \theta_0(\frac{1}{\zeta};q)\,=\,-\frac{\theta_0(\zeta;q)}{\zeta} \,.
\ee
The elliptic Gamma function has the following product representation
\begin{equation}\label{GammaeDef}
\G(\z;p,q) \=  \prod_{j,\,k=0}^{\infty}
\frac{1\,-\, \zeta^{-1} p^{j+1} q^{k+1}}{1\,-\,\zeta \,p^j\, q^k} \,.
\end{equation} 
The following properties were used in the main body of the paper
\be\label{eq:QuasiPGammas}
\G(p\z;p,q) \= \theta_0(\z;q)\G(\z;p,q)\,, \qquad  \G(q\z;p,q) \= \theta_0(\z;p)\G(\z;p,q)\,.
\ee

\section{The ambiguity of multivariate residues}\label{app:Ambiguity}

Let us focus on the irreducible boundary conditions 1\,,~\eqref{eq:AnsatzIBC1}, and one of its topological roots 
\be
\underline{\boldsymbol{z}}^{\ast}\,=\,\left\{ (-1)^{\frac{1}{3}}\,,\, (-1)^{\frac{2}{3}}\right\}\,,
\ee
the second element in~\eqref{eq:N3DiscreteSols}.

In the naive infinitesimal vicinity
\be
|\delta {x}|\,:=\,|\delta z_1|\leq \epsilon\,,\qquad |\delta{y}|\,:=\,|\delta z_2|\leq \epsilon
\ee
where
\be
\left\{z_1\,=\, (-1)^{\frac{1}{3}}\,+\,\delta x\,,\,z_2\,=\, (-1)^{\frac{2}{3}}\,+\,\delta y\right\}
\ee
the integrand looks as follows
\be
I_{\textrm{B.A}}\,=\,\begin{cases} \mathcal{O}(\delta x^0) \qquad\,\,\, \text{when $\delta x$-expansion is taken before the $\delta y$-expansion}   \\ \mathcal{O}(\delta y^0)\,, \qquad \text{when $\delta y$-expansion is taken before the $\delta x$-expansion} 
\end{cases}
\ee
and thus any residue of $\underline{\boldsymbol{z}}^{*}$ computed in this vicinity vanishes.

On the other hand, in the infinitesimal vicinity
\be
|\delta {\rho}|\,:=\,|\delta z_1|\leq \epsilon\,,\qquad | \delta {\rho}\,\theta|\,:=\,|\delta z_2|\leq \epsilon\,,
\ee
where
\be 
\theta\, := \frac{\delta{z_2}}{\delta z_1}=\, \mathcal{O}(\epsilon^0)
\ee
and
\be
\left\{z_1\,=\, (-1)^{\frac{1}{3}}\,+\,\delta\rho \,,\,z_2\,=\, (-1)^{\frac{2}{3}}\,+\,\delta\rho \,\theta\right\}\,,
\ee
the integrand looks as follows around $\delta\rho=0$
\be
I_{\textrm{B.A}}\,=\,\frac{9 \left(1+\text{i} \sqrt{3}\right)}{4 \left(\theta_--\theta \right) \left(\theta -\theta_+\right)}\,\frac{d(\delta\rho)\wedge d\theta}{\delta\rho}\,+\,\mathcal{O}(\delta\rho^0)\,\,+\, \mathcal{O}(\widetilde{q})\,.
\ee
In this vicinity there are two poles located at $\delta \rho =0$ and 
\be
\theta\,=\,\theta_- \,:=\,({-1})^{\frac{1}{3}}\,, \qquad \theta\,=\,\theta_+\,:=\,-2 ({-1})^{\frac{1}{3}} 
\ee
and their individual residues are
\be
\text{Res}_{\pm}\,:=\,\text{Res}[I_{\textrm{B.A}};\{\rho,\theta\}]\Bigl|_{\{\rho,\,\theta\}\,=\{0,\,\theta_{\pm}\}\,}\,=\,\pm\,\frac{3}{2}\,+\, \mathcal{O}(\widetilde{q})\,.
\ee
The total residue around $\underline{z}^{*}$ is an integral combination of these two residues, labelled by two integers $n_{\pm}\in \mathbb{Z}\,$. $n_+$ (resp. $n_-$) counts the times that the chosen contour surrounds the pole~$+$ (resp. the pole~${-}$) as the former shrinks to $\delta \rho =0\,$. Namely, the total residue has the form
\be
\begin{split}
\text{Res}_{\mathfrak{m}_{\underline{z}^{*}}}[I_{\textrm{B.A}};\underline{z}^{*}]&\,=\, n_+\,\text{Res}_+\,+\, n_-\,\text{Res}_{-} \\
&\,=\, (n_+ \,-\,n_-) \,\frac{3}{2} \,+\,\mathcal{O}(\widetilde{q})
\\
&\,=:\, \mathfrak{m}_{\underline{z}^{*}}\,\frac{3}{2}\,+\,\mathcal{O}(\widetilde{q})\,.
\end{split}
\ee
The choice of infinitesimal contour that we used in the main body of the text for topological roots as the one above, corresponds to $\mathfrak{m}_{\underline{z}^{*}}\,=\,n_+\,=\,1\,$. 

The same analysis extends, analogously, to all the other topological and non-topological roots.

\section{$U(2)$ index at $p\,=\,q^2 \,$: an example }
\label{sec:PneqQ}
In this case the index is (at $Y_2\neq Y_1$)
\be\label{IndexToCompare}
\begin{split}
\mathcal{I} &\,=\,1+\widetilde{q}^3 \left(Y_1+Y_2+\frac{1}{Y_2 Y_1}-1\right)\\ &+\frac{\widetilde{q}^6 \left(Y_1 Y_2
   \left(Y_2+Y_1 \left(Y_2 \left(2 Y_1^2+\left(Y_2-1\right) Y_1+Y_2 \left(2
   Y_2-1\right)-2\right)+1\right)-1\right)+2\right)}{Y_1^2
   Y_2^2}+\mathcal{O}\left(\widetilde{q}^7\right)\,.
   \end{split}
\ee
The goal is to reach $\mathcal{I}$, up to the highest possible order in powers in its small-$\widetilde{q}\,$ expansion, out of the small $\widetilde{q}$ expansion of
\be
\int_{\mathcal{C}}  dz_1\, I_{\textrm{B.A}} \,:=\, \int_{\mathcal{C}}\frac{d z_1}{z_1} \frac{\kappa I}{ (\mathcal{Q}_1-1)}\,,
\ee
where
\be
\mathcal{C}\,=\, \mathcal{C}_1-\mathcal{C}_0\,=\,\partial \mathcal{A}(p)
\ee
and
\be
\mathcal{A}(p)\,=\, \Bigl\{z_1 \in \mathbb{C}\,:\, 1\leq|z_1|<\frac{1}{p}=\frac{1}{q^2}\Bigr\}\,.
\ee
The relevant solutions to the (chosen)Bethe equation
\be
\mathcal{Q}_1=\mathcal{Q}_1(z_1)\,:=\,\mathcal{S}_{p,q}(z_1)\,=\,1\,, \qquad p\,=\,q^2\,=\,\widetilde{q}^6,
\ee
are
\be
\{z_1^\ast\}\,=\,\{\pm\frac{\sqrt{p}}{q},\pm\frac{\sqrt{p}}{q^2}\,,\,\pm\frac{\sqrt{p q}}{q^2},\pm\frac{\sqrt{p q}}{q^3}\}\,.
\ee
After the straightforward computation of
\be
\begin{split}
\int_{\mathcal{C}}\frac{d z_1}{z_1} \frac{\kappa I}{ (\mathcal{Q}_1-1)}&\,=\,\sum_{{z}^\ast_1}\text{Res}[\frac{\kappa I}{ z_1(\mathcal{Q}_1-1)};z_1^\ast] \,,\\
&\,=\,\sum_{z_1^\ast} \frac{\kappa I[z_1^\ast]}{\partial_{\log {z_1}}\mathcal{Q}_1(z^{*}_1)}
\end{split}
\ee
we indeed obtain the right-hand side of~\eqref{IndexToCompare}. 

\bibliographystyle{JHEP}

\end{document}